\documentclass[twocolumn,pra,amsmath,amssymb,floatfix,superscriptaddress,showpacs]{revtex4-1}
\usepackage{xcolor}
\usepackage{graphicx}
\usepackage{dcolumn}
\usepackage{bm}
\usepackage{times}

\begin{document}

\title{Coexistence of superconductivity and short-range double-stripe spin correlations\\ 
in Te-vapor annealed FeTe$_{1-x}$Se$_{x}$ with $x\le0.2$}

\author{Zhijun~Xu}
\affiliation{NIST Center for Neutron Research, National Institute of
Standards and Technology, Gaithersburg, Maryland 20899,
USA}\affiliation{Department of Materials Science and Engineering,
University of Maryland, College Park, Maryland 20742, USA}
\affiliation{Physics Department, University of California, Berkeley,
California 94720, USA} \affiliation{Materials Science Division,
Lawrence Berkeley National Laboratory, Berkeley, California 94720,
USA}
\author{J.~A.~Schneeloch}
\affiliation{Condensed Matter Physics and Materials Science
Division, Brookhaven National Laboratory, Upton, New York 11973,
USA} \affiliation{Department of Physics, Stony Brook University,
Stony Brook, New York 11794, USA}
\author{Ming~Yi}
\affiliation{Physics Department, University of California, Berkeley,
California 94720, USA} \affiliation{Materials Science Division,
Lawrence Berkeley National Laboratory, Berkeley, California 94720,
USA}
\author{Yang~Zhao}
\affiliation{NIST Center for Neutron Research, National Institute of
Standards and Technology, Gaithersburg, Maryland 20899,
USA}\affiliation{Department of Materials Science and Engineering,
University of Maryland, College Park, Maryland 20742, USA}
\author{Masaaki~Matsuda}
\author{D.~M.~Pajerowski}
\author{Songxue~Chi}
\affiliation{Neutron Scattering Division, Oak Ridge National
Laboratory, Oak Ridge, Tennessee 37831, USA}
\author{R.~J.~Birgeneau}
\affiliation{Physics Department, University of California, Berkeley,
California 94720, USA} \affiliation{Materials Science Division,
Lawrence Berkeley National Laboratory, Berkeley, California 94720,
USA}
\author{Genda~Gu}
\author{J.~M.~Tranquada}
\affiliation{Condensed Matter Physics and Materials Science
Division, Brookhaven National Laboratory, Upton, New York 11973,
USA}
\author{Guangyong~Xu}
\affiliation{NIST Center for Neutron Research, National Institute of
Standards and Technology, Gaithersburg, Maryland 20899, USA}
\affiliation{Condensed Matter Physics and Materials Science
Division, Brookhaven National Laboratory, Upton, New York 11973,
USA}
\date{\today}

\begin{abstract}
In as-grown bulk crystals of Fe$_{1+y}$Te$_{1-x}$Se$_{x}$ with $x\lesssim0.3$, excess Fe ($y>0$) is inevitable and correlates with a suppression of superconductivity.  At the same time, there remains the question as to whether the character of the antiferromagnetic correlations associated with the enhanced anion height above the Fe planes in Te-rich samples is compatible with superconductivity.  To test this, we have annealed as-grown crystals with $x=0.1$ and 0.2 in Te vapor, effectively reducing the excess Fe and inducing bulk superconductivity.  Inelastic neutron scattering measurements reveal low-energy magnetic excitations consistent with short-range correlations of the double-stripe type; nevertheless, cooling into the superconducting state results in a spin gap and a spin resonance, with the extra signal in the resonance being short-range with a mixed single-stripe/double-stripe character, which is different than other iron-based superconductors.  The mixed magnetic character of these superconducting samples does not appear to be trivially explainable by inhomogeneity.
%
\end{abstract}


\maketitle

\section{Introduction}

The interplay between superconductivity and magnetism  is still
one of the main topics in the field of high-temperature
superconductivity \cite{Mazin10,Lynn2009,pagl10,scal12a,tran14}. While commensurate
antiferromagnetic (AF) order appears to compete with superconductivity, 
magnetic excitations are widely believed to be important in mediating electron pairing in many
high-$T_c$ superconductors \cite{Tranquada2004,Gxu2009,Rossat1991,Bourges1996,Dai2000,Fong1999,Hayden2004,Vignolle2007,Hinkov2007np,pagl10,scal12a,tran14,lums10r}.  One of the most important signatures of the coupling between magnetic excitations and superconductivity is the ``spin resonance'', where magnetic intensity detected by neutron scattering at the resonance energy exhibits a sharp enhancement when the system is cooled into the superconducting (SC) state \cite{Christianson2008,Christianson2013,Lumsden2009prl,Chis2009prl,Inosov2010nf,Qiu2009,Wen2010H,yu09}.

In many Fe-based superconductors (FBS), such as the `122' \cite{Christianson2008,Lumsden2009prl,Chis2009prl,Lis2009prb,Inosov2010nf},`1111' \cite{Wakimoto2010} and `111' families \cite{Zhang2013,Zhang2015}, the magnetic order in the parent compound~\cite{Dai2015} corresponds to the stripe antiferromagnet (SAF), characterized by the in-plane wave vector $Q_{\rm SAF}=(0.5,0.5)$, and
the spin-resonance in the SC compositions appears at the same location in momentum space.
This is not the case in FeTe$_{1-x}$Se$_{x}$, which is known as the `11' system~\cite{Bao2009,Lynn2009,Lumsden2010nf,zxu2010fetese1,Xu2016}. Here the parent compound
Fe$_{1+y}$Te exhibits long-range AF order made up of double stripes of parallel spins within each Fe layer.  Based on a crystallographic unit cell containing two Fe atoms, the in-plane component of this double-stripe antiferromagnetic (DSAF) order is characterized by the wave vector $Q_{\rm DSAF}=(0.5,0)$, with spin-wave type magnetic excitations emerging from $Q_{\rm DSAF}$~\cite{Lipscombe2011,Lumsden2010nf,Zaliznyak2011}. When sufficient Se is substituted to yield bulk superconductivity, a spin resonance is observed, but it occurs at $Q_{\rm SAF}$ as in the other FBS families \cite{Qiu2009,Lumsden2010nf,Lee2010,Xu2016}. The magnetic excitations tend to disperse out from $Q_{\rm SAF}$ in the transverse directions, with the bottom of the dispersion being around 5~meV, and the spin resonance occurs around $\hbar\omega=6.5$~meV.   A unique feature of FeTe$_{1-x}$Se$_x$ is that the character of the low-energy magnetic excitations changes dramatically with temperature \cite{Xu2016,Xu2017}.  Well above the superconducting critical temperature, $T_c$, the low-energy magnetic excitations shift away from $Q_{\rm SAF}$ and instead develop the signature of short-range correlations associated with a local DSAF modulation. 

\begin{figure}[t]
	\includegraphics[width=0.9\linewidth]{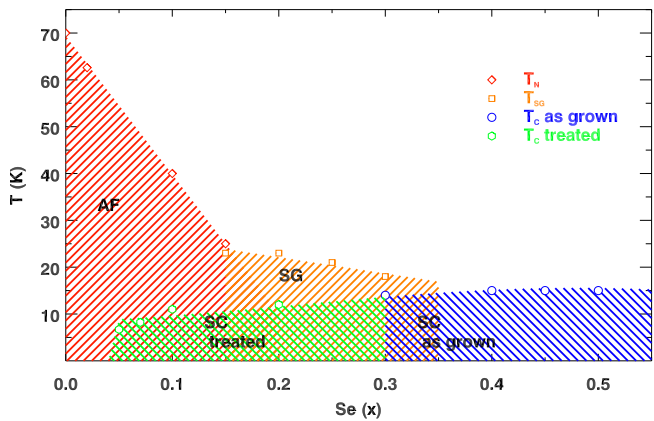}
	\caption{(Color online) Phase diagram of FeTe$_{1-x}$Se$_{x}$ as function of Se content ($x$) and temperature ($T$). The red circles represent the N\'eel temperature ($T_{N}$); blue circles represent the as-grown samples' superconducting onset temperature $T_{c}$; purple circles represent the superconducting onset temperature in the treated samples. Data from Refs.~\onlinecite{Katayama2010,Dong2011} are included here. } \label{fig:6}
\end{figure}

As shown in Fig.~\ref{fig:6}, the long-range DSAF order in Fe$_{1+y}$Te$_{1-x}$Se$_x$  disappears at $x\approx0.1$; it is associated with an orthorhombic lattice distortion  that disappears at the same Se concentration \cite{mart10b}.  In as-grown crystals, bulk superconductivity appears for $x\gtrsim0.3$ \cite{Katayama2010,liu10}, while glassy, short-range DSAF order coexists with weak, inhomogeneous superconductivity for $0.1<x<0.3$.
Studies deliberately varying the concentration $y$ of excess Fe have shown that  the excess 
is correlated with the suppression of superconductivity, especially in this intermediate range of $x$ \cite{Bendele2010,rodr11}.  By reducing the excess Fe in such samples, one can drive the system towards SC \cite{Bendele2010,rodr11,Dong2011}.  
There are several different annealing methods available for this purpose, including annealing in air, oxygen, Se, Te and S vapor \cite{Koshika2013,Dong2011,Sun2013}.  In this work, we use Te
vapor~\cite{Koshika2013}, which avoids the introduction of extra elements
such as oxygen while maintaining a high Te concentration.

In this paper, we report a systematic study of the magnetic correlations in single crystals of Fe$_{1+y}$Te$_{1-x}$Se$_{x}$ with $x=0.1$ and 0.2 that have been annealed in Te vapor for sufficient time to yield bulk superconductivity.
Our neutron scattering measurements reveal low-energy magnetic excitations with a {\bf Q} dependence characteristic of short-range DSAF correlations, as seen previously in FeTe$_{0.87}$S$_{0.13}$ \cite{zali15}.  The new feature here is that we also observed a spin gap and resonance for $T<T_c$.  The increase in signal associated with the resonance has a {\bf Q} dependence that  appears to mix the characteristics of SAF and DSAF correlations, which, in turn, is different than the pure SAF spin correlations observed at low temperature in other SC FeTe$_{1-x}$Se$_x$ samples \cite{Qiu2009,Lumsden2010nf,Lee2010,Xu2016}.  This provides an interesting test case for theoretical models that connect the magnetism and superconductivity.


\section{Experimental Details}

\begin{figure}[t]
\includegraphics[width=0.9\linewidth]{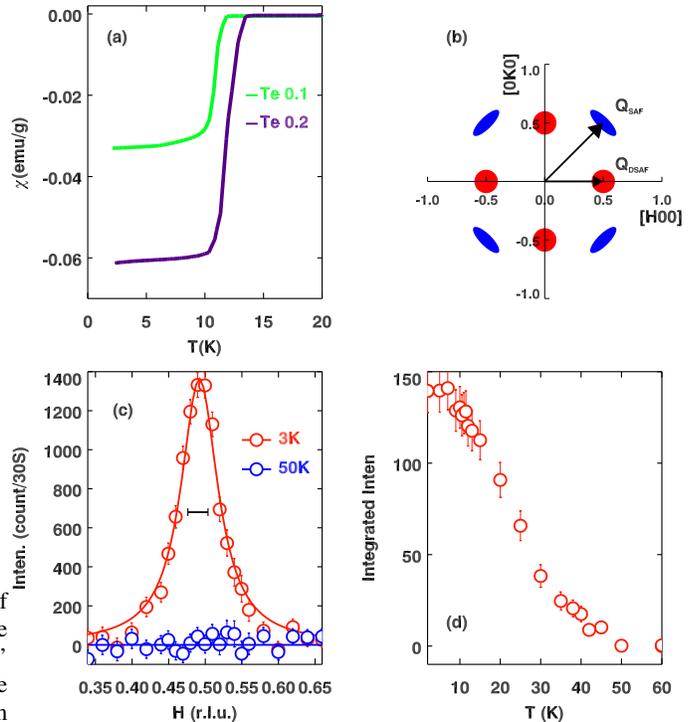}
\caption{(Color online) (a) Zero-field-cooled magnetization measurements by SQUID
with a 10~Oe field perpendicular to the $a$-$b$ plane for all
samples: $x = 0.1$ (green solid line) and $x = 0.2$ (purple solid line).
(b) Diagram of reciprocal space indicating the characteristic wave vectors ${\bf Q}_{\rm SAF}$ and ${\bf Q}_{\rm DSAF}$.
(c) Elastic neutron-scattering measurements performed on $x = 0.1$
sample around magnetic order peak at $(0.5,0,0.5)$ measured on BT-7.
Intensity profiles along [100] direction ($H$ scans) at temperatures
below ($T=3$~K, red) and above $T_{N}$ (50~K, blue).  The horizontal (black) bar represents the $H$ resolution.  (d) The integrated magnetic peak intensity (from fitted Gaussian peak intensity) vs temperature. The error bars represent the square root of the number
of counts.} \label{fig:1}
\end{figure}

\begin{table}
	\caption{List of the Fe$_{1+y}$Se$_x$Te$_{1-x}$
		samples used in our measurements, with their Fe composition before and after annealing in Te vapor measured by EDX spectroscopy, and the superconducting transition temperature, $T_{c}$, obtained from the magnetic susceptibility measurements in Fig.~\ref{fig:1}(a). }
	\begin{ruledtabular}
		\begin{tabular}{cccccc}
			Sample & As-grown & Annealed & $T_{c}$   \\
			&      &    &   (K)       \\
			\hline
			x = 0.1 & y=0.025 & y=-0.027 & 12  \\
			x = 0.2 & y=0.096 & y=0.045 & 13   \\
		\end{tabular}
	\end{ruledtabular}
	\label{tab:1}
\end{table}

The single-crystal samples used in this experiment were
grown by a unidirectional solidification method~\cite{JWen2011} at
Brookhaven National Laboratory. The as-grown single crystals, which
contained excess Fe and were not superconducting ~\cite{Katayama2010},
were annealed at 400 $^{\circ}$C for 10 days in Tellurium (Te)
vapor~\cite{Koshika2013}.  The Fe excess, $y$, before and after annealing was measured by Energy-Dispersive X-ray (EDX) spectroscopy; the results listed in Table~I indicate that the Te-vapor annealing caused a substantial reduction in $y$.  The
bulk susceptibilities, measured with a superconducting quantum
interference device (SQUID) magnetometer, are shown in
Fig.~\ref{fig:1}(a).  They demonstrate a bulk superconducting response for each sample, though less than 100\%\ shielding fraction.

Neutron scattering experiments were carried out
on the triple-axis spectrometers BT-7~\cite{Lynn2012} at NIST Center
for Neutron Research (NCNR) and HB-1 located at the High Flux
Isotope Reactor (HFIR) at Oak Ridge National Laboratory (ORNL). We
used beam collimations of open-$80'$-S-$80'$-$120'$ (S = sample)
with fixed final energy of 14.7~meV and two pyrolytic graphite (PG)
filters after the sample to reduce higher-order neutrons at BT-7 and
$48'$-$80'$-S-$80'$-$120'$ with the same fixed final energy and one
PG filter after the sample at HB-1. Except for the elastic
scattering measurements in Fig.~\ref{fig:1}, which were performed in the
$(H0L)$ scattering plane, all inelastic scattering measurements were
performed in the $(HK0)$ scattering plane. The lattice constants for
these samples are $a = b \approx 3.8$~\AA, and $c \approx 6.1$~\AA,
using a unit cell containing two Fe atoms. The wave vectors are specified in
reciprocal lattice units (r.l.u.) of $(a^*, b^*, c^*) = (2\pi/a,
2\pi/b, 2\pi/c)$.

\section{Results}

We have performed a series of neutron scattering measurements on the
Te-vapor annealed superconducting samples of FeTe$_{1-x}$Se$_x$. 
We started with elastic measurements to test for static magnetic order in the $x=0.1$
sample. In Fig.~\ref{fig:1} (c), we plot $H$
scans through the $ {\bold Q}_{\rm AF} \approx (0.5, 0, 0.5)$ wave vector at $T =3$~K and
50~K.  The magnetic peak observed at low temperature is broader than experimental resolution, and the peak center is slightly incommensurate, consistent with previous results \cite{Katayama2010,Wen2009}. 
The integrated intensity of this peak, shown in Fig.~\ref{fig:1}(d), 
gradually decreases upon heating and  disappears around 40~K, consistent with susceptibility measurements on air-annealed superconducting crystals with similar $x$ \cite{Dong2011}.   As we will see next, the low-energy inelastic magnetic scattering bears no simple connection to these elastic peaks, and hence we believe that the static order occurs in a minority of the sample volume that is likely segregated from the superconducting regions.  We note that a recent scanning tunneling microscopy study on an $x=0.1$ sample found evidence for local coexistence of AF order and  pairing gaps 
\cite{alur17}; however, that sample did not exhibit the degree of bulk  superconducting order found in our crystal.


\begin{figure}[t]
\hskip20pt\includegraphics[width=0.9\linewidth]{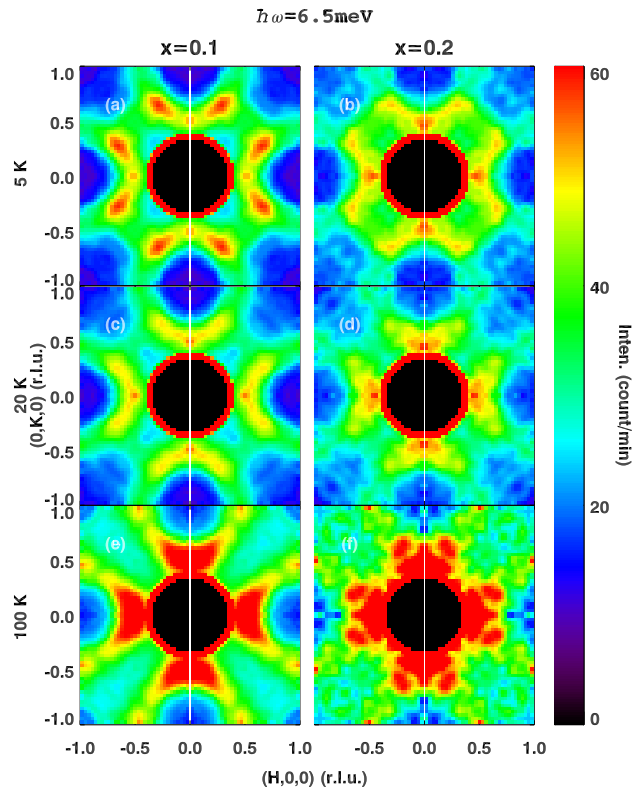}
\caption{(Color online) Contour intensity maps of magnetic neutron
scattering intensity measured on HB-1 in $(HK0)$ plane at energy
transfer $\hbar\omega=6.5$~meV. The maps are plotted for the $x = 0.1$
(left column) and $x = 0.2$ (right column) samples at sample
temperatures: (a), (b) 5~K, (c), (d) 20~K and (e), (f) 100~K.
The data have been folded from the first quadrant ($H > 0$, $K> 0$).
Intensity scale is the same in all panels and the data have been
smoothed.} \label{fig:2}
\end{figure}

Next, we consider measurements of the low-energy magnetic excitations.  
Figure~\ref{fig:2} shows color contour plots of spin excitations measured in the $(HK0)$
plane at an energy of 6.5~meV, which
corresponds to the spin-resonance energy at optimal doping in this compound~\cite{Qiu2009}. Panels in the left column show data from the $x=0.1$ sample at temperatures of 5, 20, and 100~K.  The data in the right column for $x=0.2$ correspond to lower counting statistics, but are qualitatively similar to those for $x=0.1$.  At $T = 5$~K, well below $T_c$, the data are quite different from the simple commensurate ellipse shape at $ {\bf Q}=(0.5, 0.5)$ seen previously for optimal doping \cite{Qiu2009,Lee2010,Xu2016}.  Instead, they closely resemble the model of short-range double-stripe correlations proposed in a study of FeTe$_{0.87}$S$_{0.13}$ by Zaliznyak {\it et al.} \cite{zali15}.   Note that the intensity pattern associated with the short-range correlations is not characterized by a well-defined wave vector; rather, it involves a distribution of spectral weight that is broad in {\bf Q} and that, in the vicinity of ${\bf Q}_{\rm SAF}$, appears incommensurate.

%

\begin{figure}[t]
\hskip20pt\includegraphics[width=0.9\linewidth]{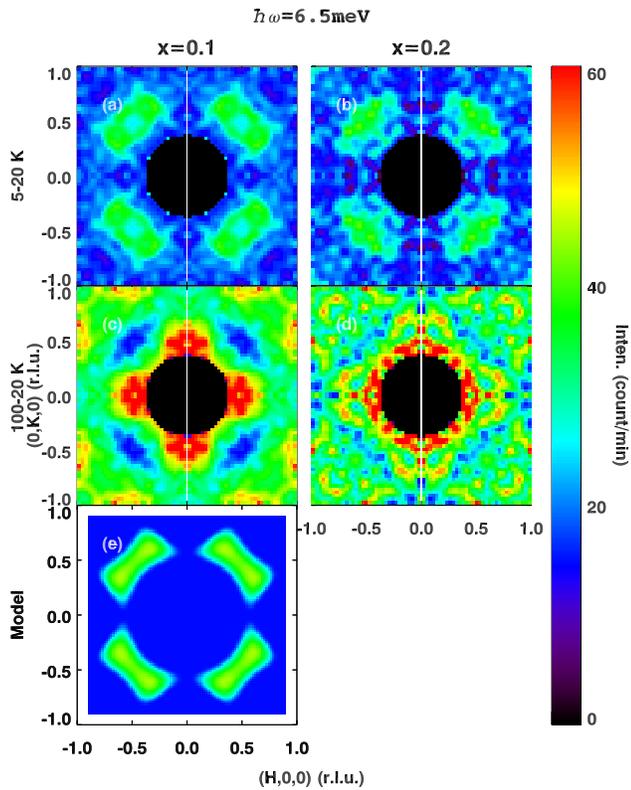}
\caption{(Color online) Contour intensity maps of temperature
difference of magnetic neutron scattering intensity measured on HB-1
in $(HK0)$ plane at energy transfer $\hbar\omega=6.5$~meV. The maps
are plotted for the $x = 0.1$ (left column) and $x = 0.2$ (right column)
samples at temperature differences of: (a), (b) $5-20$~K, and (c), (d)
$100-20$~K. The data have been folded from first quadrant ($H>0$, $K>0$). (e) Intensity calculated based on the same UDUD  spin-plaquette model described in Ref.~\onlinecite{Xu2016,Zaliznyak2015}, with the volume ratio of interplaquette correlation being 25\% SAF and 75\% DSAF. Intensity scale is the same in all panels and the data
have been smoothed.} \label{fig:3}
\end{figure}

The change in the scattering pattern on warming 
across $T_c$ is subtle, but the changes are larger when the temperature is increased to 100~K.  To get a better view of the changes, temperature differences are plotted in Fig.~\ref{fig:3}.  The difference between 5 and 20 K for the $x=0.1$ sample shown in Fig.~\ref{fig:3} (in contrast to the absolute signal at 5 K) is 
similar to measurements of the resonance in optimally-superconducting FeTe$_{1-x}$Se$_x$ \cite{Qiu2009,Lee2010,Xu2016}. However,  the intensity  maxima are not located at the commensurate $(0.5,0.5)$ positions but slightly further out in the transverse directions. One can see that the difference, which is indeed the Q-distribution of the spin resonance, appears to be highly consistent with a model calculation [Fig.~\ref{fig:3} (e)] based on the same UDUD spin plaquette model described in Ref.~\onlinecite{Xu2016,Zaliznyak2015}, with the volume ratio of interplaquette correlation being 25\% SAF and 75\% DSAF. On the other hand, the difference between 100 and 20 K bears the signature of ferromagnetic plaquettes with short-range antiferromagnetic correlations, as previously discussed for FeTe$_{0.87}$S$_{0.13}$ \cite{zali15}, where such a component was also found to be enhanced with increasing temperature.
The data from the $x=0.2$ sample are less informative but are qualitatively in agreement with the $x=0.1$ data.

\begin{figure}[t]
\hskip20pt\includegraphics[width=0.9\linewidth]{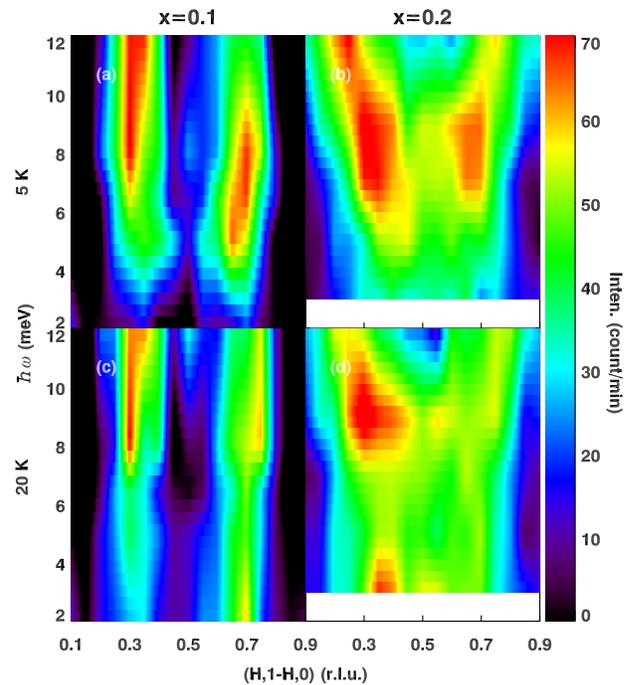}
\caption{(Color online) Contour intensity maps of magnetic neutron
scattering intensity in energy-momentum space along the transverse
direction. The maps are plotted for the $x = 0.1$ (left column,
measured on BT-7) and $x = 0.2$ (right column, measured on HB-1)
samples at sample temperatures: (a), (b) 5~K and (c), (d) 20~K. The
data have been smoothed.} \label{fig:4}
\end{figure}

To characterize the energy dispersion in the vicinity of the resonance, we plot in Fig.~\ref{fig:4} the energy dependence of the magnetic scattering along the transverse direction, ${\bf Q}=(H,1-H,0)$,  around $H=0.5$. 
As one can see, the low-energy dispersion in the $x=0.1$ sample takes the form of two vertical columns; in the case of $x=0.2$, the commensurate region between the columns has begun to fill in.  In both cases, a comparison of the data at 5 and 20 K clearly reveals the opening of a spin gap below 5 meV and the intensity enhancement of the resonance above that.


\begin{figure}[t]
\hskip20pt\includegraphics[width=0.8\linewidth]{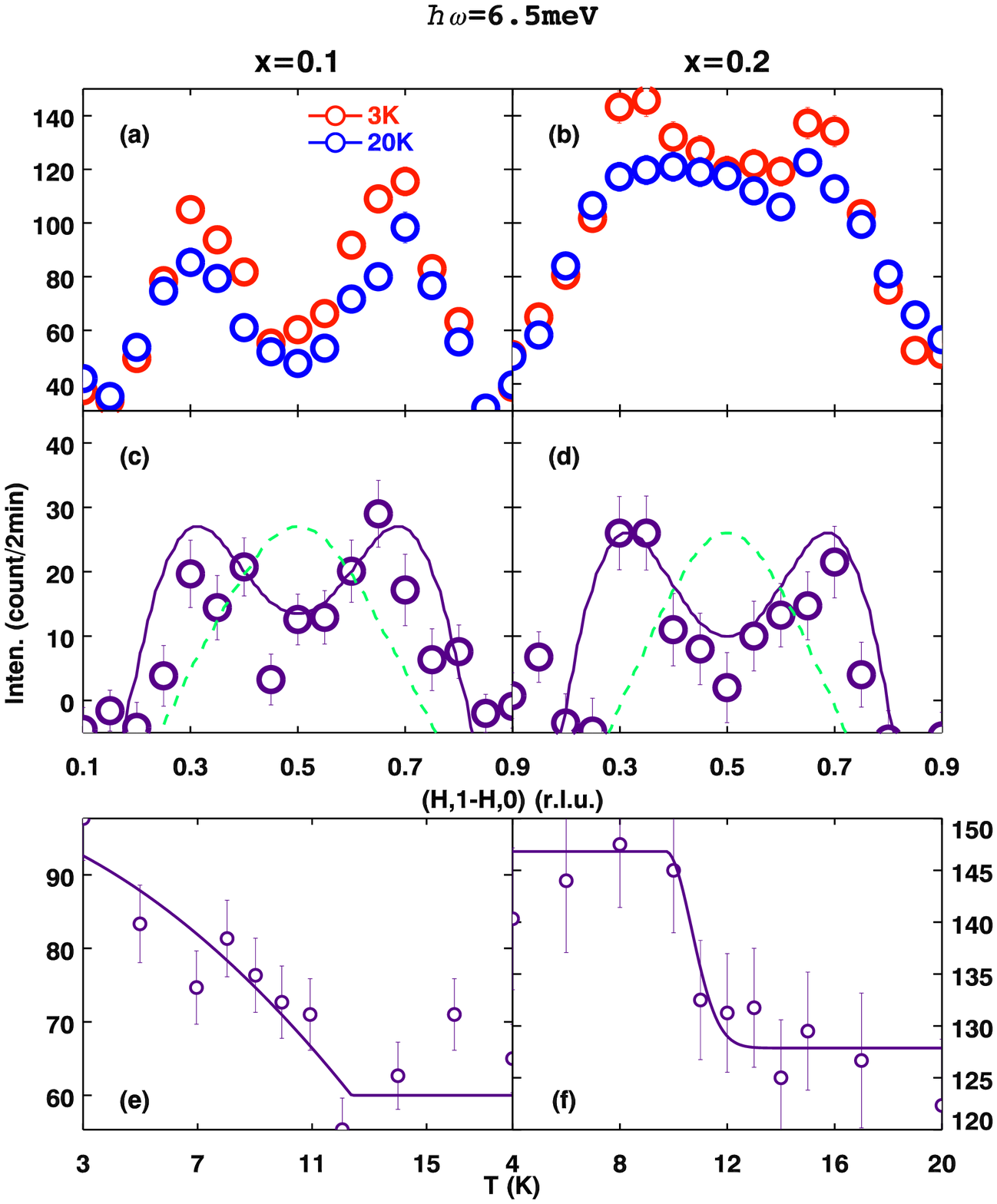}
\caption{(Color online) Constant energy scans of magnetic scattering
intensity along the transverse direction at excitation energy
6.5~meV for the (a) $x = 0.1$ and (b) $x = 0.2$ samples at sample
temperatures: 3~K (red circles) and 20 K (blue circles). The wave
vector dependence of the spin resonance from the temperature difference 
$3-20$~K is plotted in (c) $x = 0.1$ and (d) $x = 0.2$. The purple lines are model calculation based on the same UDUD spin plaquette model described in Ref.~\onlinecite{Xu2016,Zaliznyak2015}, with the volume ratio of interplaquette correlation being 25\% SAF and 75\% DSAF in (c) and 20\% SAF and 80\% DSAF in (d). The green dashed lines are a similar model calculation based on 100\% SAF correlations. (e) and (f) show the temperature dependence of the spin resonance from
peak intensities at $(0.6, 0.4, 0)$ at 6.5~meV for respective samples.
The error bars represent the
square root of the number of counts. } \label{fig:5}
\end{figure}

For a more detailed look at the resonance, Fig.~\ref{fig:5} shows constant-energy scans along the
transverse direction at 6.5~meV obtained at 3 and 20~K. 
By subtracting the 20~K data from the 3~K data, the {\bf Q} dependence of the intensity enhancements is displayed in Fig.~\ref{fig:5}(c) and (d).  The response is strongly peaked at incommensurate positions with incommensurability $\sim 0.08$. One can clearly see the discrepancy between  model calculations based on a phase with 100\% SAF correlations [green dashed lines in panels (c) and (d)] and the measured $q$-distribution of the resonance. Instead, only when we consider a phase with mixed SAF and DSAF correlations, can the incommensurate response be reproduced. As shown in the insets of Fig.~\ref{fig:5}(c) and
(d), the spin resonance
intensity starts to rise on cooling below 12~K in the $x = 0.1$ sample and
below 13~K in the $x = 0.2$ sample, consistent with the $T_{c}$ values obtained from the susceptibility 
measurements in Fig.~\ref{fig:1}(a). 

\section{Discussion}

In our Te-rich crystals of FeTe$_{1-x}$Se$_x$, we have observed low-energy magnetic excitations consistent with short-range double-stripe spin correlations coexisting with bulk superconductivity.  In evaluating this coexistence, we must certainly take account of inhomogeneity.  For example, we also see elastic magnetic scattering consistent with intermediate-range DSAF order, which we expect is in a limited volume of each sample, spatially segregated from the superconductivity.  It is possible that the Te-vapor annealing was not done for a sufficiently long time to homogeneously modify all regions of our large crystals.  Of course, there is always the intrinsic inhomogeneity associated with the difference in local Fe-Te and Fe-Se bond lengths \cite{louc10} and the tendency to spatial segregation \cite{hu11}.  The key observation, however, is that the magnetic scattering changes across $T_c$, developing both a spin gap and resonance peak. The resonance intensity, which is not sensitive to any possible  nonsuperconducting portion of the sample, appears at incommensurate positions, slightly away from (0.5,0.5). Measuring the resonance provides a direct probe  of 
the SC portion of the sample(s) even with  a nonsuperconducting 
portion present. Our results imply that the spin correlations from the SC portion of our Te-vapor treated samples exhibit a mixed DSAF and SAF character, distinct from 
the typical behavior in SC FeTe$_{1-x}$Se$_x$ systems at low temperature. This provides a clear indication of  superconductivity developing locally within regions where the spin correlations have substantial DSAF character. 

The low-temperature two-column dispersion along $(H,1-H,0)$ has been observed previously, but in association with the suppression of superconductivity in Cu-doped FeTe$_{0.5}$Se$_{0.5}$ \cite{xu12a}.  The same dispersion is also seen at high temperatures in samples with optimal superconductivity \cite{xu12a,tsyr12,Xu2016}.  It was previously pointed out that the thermal evolution of the spin correlations is connected to the change in the tetrahedral bond angles \cite{Xu2016,Xu2017} which results in changes in hybridization between Fe $3d$ orbitals and ligand $p$ orbitals \cite{yin11b}.  Of course, the average bond angles also change with Se concentration.  It appears that we can roughly correlate the pattern of low-energy magnetic scattering in reciprocal space with the ratio of lattice parameters, $a/c$.   

The interesting point is that, while the {\bf Q} dependence of the low-energy magnetic scattering may vary significantly with composition, the resonance always appears in the vicinity of $(0.5,0.5,0)$.  The general pattern of the magnetic scattering in our samples is not compatible with simple Fermi-surface nesting arguments \cite{Yin2010}; nevertheless, the wave vectors at which the resonance occurs connect Fermi surface pockets about the $\Gamma$ and $M$ points of the Brillouin zone where the superconducting gap appears \cite{sark17,maie09,scal12a}.  The magnetic excitations certainly appear to interact with the superconducting electrons; however, the general relationship between the magnetism and superconductivity in these samples is less clear.  Analyzing this relationship, taking account of the present results, could lead to new insights into the pairing mechanism in iron chalcogenides.

\section{Summary}

We have used Te-vapor annealing to induce bulk superconductivity in crystals of Fe$_{1+y}$Te$_{1-x}$Se$_x$ with $x=0.1$ and 0.2.  Neutron scattering measurements reveal low-energy magnetic excitations with a wave vector dependence characteristic of short-range DSAF spin correlations.  While the presence of such correlations at low temperature has previously been associated with suppressed superconductivity, we find that the excitations in the vicinity of, but not exactly at, $(0.5,0.5,0)$ develop a spin gap and resonance peak.  Thus, it appears that superconductivity can coexist with magnetic correlations different from the common stripe form.  These results provide an interesting test case for understanding the relationship between magnetism and superconductivity in the iron chalcogenides.


\section{Acknowledgments}
The work at Brookhaven National Laboratory and Lawrence Berkeley
National Laboratory was supported by the Office of Basic Energy
Sciences (BES), Division of Materials Science and Engineering, U.S.
Department of Energy (DOE), under Contract Nos.\  DE-SC0012704
and DE-AC02-05CH1123, respectively. A portion of this research used resources at the High Flux Isotope Reactor, a DOE Office of Science User Facility operated by the Oak Ridge National Laboratory.


\begin{thebibliography}{58}%
	\makeatletter
	\providecommand \@ifxundefined [1]{%
		\@ifx{#1\undefined}
	}%
	\providecommand \@ifnum [1]{%
		\ifnum #1\expandafter \@firstoftwo
		\else \expandafter \@secondoftwo
		\fi
	}%
	\providecommand \@ifx [1]{%
		\ifx #1\expandafter \@firstoftwo
		\else \expandafter \@secondoftwo
		\fi
	}%
	\providecommand \natexlab [1]{#1}%
	\providecommand \enquote  [1]{``#1''}%
	\providecommand \bibnamefont  [1]{#1}%
	\providecommand \bibfnamefont [1]{#1}%
	\providecommand \citenamefont [1]{#1}%
	\providecommand \href@noop [0]{\@secondoftwo}%
	\providecommand \href [0]{\begingroup \@sanitize@url \@href}%
	\providecommand \@href[1]{\@@startlink{#1}\@@href}%
	\providecommand \@@href[1]{\endgroup#1\@@endlink}%
	\providecommand \@sanitize@url [0]{\catcode `\\12\catcode `\$12\catcode
		`\&12\catcode `\#12\catcode `\^12\catcode `\_12\catcode `\%12\relax}%
	\providecommand \@@startlink[1]{}%
	\providecommand \@@endlink[0]{}%
	\providecommand \url  [0]{\begingroup\@sanitize@url \@url }%
	\providecommand \@url [1]{\endgroup\@href {#1}{\urlprefix }}%
	\providecommand \urlprefix  [0]{URL }%
	\providecommand \Eprint [0]{\href }%
	\providecommand \doibase [0]{http://dx.doi.org/}%
	\providecommand \selectlanguage [0]{\@gobble}%
	\providecommand \bibinfo  [0]{\@secondoftwo}%
	\providecommand \bibfield  [0]{\@secondoftwo}%
	\providecommand \translation [1]{[#1]}%
	\providecommand \BibitemOpen [0]{}%
	\providecommand \bibitemStop [0]{}%
	\providecommand \bibitemNoStop [0]{.\EOS\space}%
	\providecommand \EOS [0]{\spacefactor3000\relax}%
	\providecommand \BibitemShut  [1]{\csname bibitem#1\endcsname}%
	\let\auto@bib@innerbib\@empty
	\bibitem [{\citenamefont {Mazin}(2010)}]{Mazin10}%
	\BibitemOpen
	\bibfield  {author} {\bibinfo {author} {\bibfnamefont {I.~I.}\ \bibnamefont
			{Mazin}},\ }\href {\doibase 10.1038/nature08914} {\bibfield  {journal}
		{\bibinfo  {journal} {Nature}\ }\textbf {\bibinfo {volume} {464}},\ \bibinfo
		{pages} {183} (\bibinfo {year} {2010})}\BibitemShut {NoStop}%
	\bibitem [{\citenamefont {Lynn}\ and\ \citenamefont {Dai}(2009)}]{Lynn2009}%
	\BibitemOpen
	\bibfield  {author} {\bibinfo {author} {\bibfnamefont {J.~W.}\ \bibnamefont
			{Lynn}}\ and\ \bibinfo {author} {\bibfnamefont {P.}~\bibnamefont {Dai}},\
	}\href@noop {} {\bibfield  {journal} {\bibinfo  {journal} {Physica C}\
		}\textbf {\bibinfo {volume} {469}},\ \bibinfo {pages} {469} (\bibinfo {year}
		{2009})}\BibitemShut {NoStop}%
	\bibitem [{\citenamefont {Paglione}\ and\ \citenamefont
		{Greene}(2010)}]{pagl10}%
	\BibitemOpen
	\bibfield  {author} {\bibinfo {author} {\bibfnamefont {J.}~\bibnamefont
			{Paglione}}\ and\ \bibinfo {author} {\bibfnamefont {R.~L.}\ \bibnamefont
			{Greene}},\ }\href@noop {} {\bibfield  {journal} {\bibinfo  {journal} {Nat.
				Phys.}\ }\textbf {\bibinfo {volume} {6}},\ \bibinfo {pages} {645} (\bibinfo
		{year} {2010})}\BibitemShut {NoStop}%
	\bibitem [{\citenamefont {Scalapino}(2012)}]{scal12a}%
	\BibitemOpen
	\bibfield  {author} {\bibinfo {author} {\bibfnamefont {D.~J.}\ \bibnamefont
			{Scalapino}},\ }\href@noop {} {\bibfield  {journal} {\bibinfo  {journal}
			{Rev. Mod. Phys.}\ }\textbf {\bibinfo {volume} {84}},\ \bibinfo {pages}
		{1383} (\bibinfo {year} {2012})}\BibitemShut {NoStop}%
	\bibitem [{\citenamefont {Tranquada}\ \emph {et~al.}(2014)\citenamefont
		{Tranquada}, \citenamefont {Xu},\ and\ \citenamefont {Zaliznyak}}]{tran14}%
	\BibitemOpen
	\bibfield  {author} {\bibinfo {author} {\bibfnamefont {J.~M.}\ \bibnamefont
			{Tranquada}}, \bibinfo {author} {\bibfnamefont {G.}~\bibnamefont {Xu}}, \
		and\ \bibinfo {author} {\bibfnamefont {I.~A.}\ \bibnamefont {Zaliznyak}},\
	}\href@noop {} {\bibfield  {journal} {\bibinfo  {journal} {J. Magn. Magn.
				Mat.}\ }\textbf {\bibinfo {volume} {350}},\ \bibinfo {pages} {148} (\bibinfo
		{year} {2014})}\BibitemShut {NoStop}%
	\bibitem [{\citenamefont {Tranquada}\ \emph {et~al.}(2004)\citenamefont
		{Tranquada}, \citenamefont {Woo}, \citenamefont {Perring}, \citenamefont
		{Goka}, \citenamefont {Gu}, \citenamefont {Xu}, \citenamefont {Fujita},\ and\
		\citenamefont {Yamada}}]{Tranquada2004}%
	\BibitemOpen
	\bibfield  {author} {\bibinfo {author} {\bibfnamefont {J.~M.}\ \bibnamefont
			{Tranquada}}, \bibinfo {author} {\bibfnamefont {H.}~\bibnamefont {Woo}},
		\bibinfo {author} {\bibfnamefont {T.~G.}\ \bibnamefont {Perring}}, \bibinfo
		{author} {\bibfnamefont {H.}~\bibnamefont {Goka}}, \bibinfo {author}
		{\bibfnamefont {G.~D.}\ \bibnamefont {Gu}}, \bibinfo {author} {\bibfnamefont
			{G.}~\bibnamefont {Xu}}, \bibinfo {author} {\bibfnamefont {M.}~\bibnamefont
			{Fujita}}, \ and\ \bibinfo {author} {\bibfnamefont {K.}~\bibnamefont
			{Yamada}},\ }\href {\doibase 10.1038/nature02574} {\bibfield  {journal}
		{\bibinfo  {journal} {Nature}\ }\textbf {\bibinfo {volume} {429}},\ \bibinfo
		{pages} {534} (\bibinfo {year} {2004})}\BibitemShut {NoStop}%
	\bibitem [{\citenamefont {Xu}\ \emph {et~al.}(2009)\citenamefont {Xu},
		\citenamefont {Gu}, \citenamefont {Hucker}, \citenamefont {Fauque},
		\citenamefont {Perring}, \citenamefont {Regnault},\ and\ \citenamefont
		{Tranquada}}]{Gxu2009}%
	\BibitemOpen
	\bibfield  {author} {\bibinfo {author} {\bibfnamefont {G.}~\bibnamefont
			{Xu}}, \bibinfo {author} {\bibfnamefont {G.~D.}\ \bibnamefont {Gu}}, \bibinfo
		{author} {\bibfnamefont {M.}~\bibnamefont {Hucker}}, \bibinfo {author}
		{\bibfnamefont {B.}~\bibnamefont {Fauque}}, \bibinfo {author} {\bibfnamefont
			{T.~G.}\ \bibnamefont {Perring}}, \bibinfo {author} {\bibfnamefont {L.~P.}\
			\bibnamefont {Regnault}}, \ and\ \bibinfo {author} {\bibfnamefont {J.~M.}\
			\bibnamefont {Tranquada}},\ }\href {\doibase 10.1038/nphys1360} {\bibfield
		{journal} {\bibinfo  {journal} {Nat. Phys.}\ }\textbf {\bibinfo {volume}
			{5}},\ \bibinfo {pages} {642} (\bibinfo {year} {2009})}\BibitemShut {NoStop}%
	\bibitem [{\citenamefont {Rossat-Mignod}\ \emph {et~al.}(1991)\citenamefont
		{Rossat-Mignod}, \citenamefont {Regnault}, \citenamefont {Vettier},
		\citenamefont {Bourges}, \citenamefont {Burlet}, \citenamefont {Bossy},
		\citenamefont {Henry},\ and\ \citenamefont {Lapertot}}]{Rossat1991}%
	\BibitemOpen
	\bibfield  {author} {\bibinfo {author} {\bibfnamefont {J.}~\bibnamefont
			{Rossat-Mignod}}, \bibinfo {author} {\bibfnamefont {L.~P.}\ \bibnamefont
			{Regnault}}, \bibinfo {author} {\bibfnamefont {C.}~\bibnamefont {Vettier}},
		\bibinfo {author} {\bibfnamefont {P.}~\bibnamefont {Bourges}}, \bibinfo
		{author} {\bibfnamefont {P.}~\bibnamefont {Burlet}}, \bibinfo {author}
		{\bibfnamefont {J.}~\bibnamefont {Bossy}}, \bibinfo {author} {\bibfnamefont
			{J.~Y.}\ \bibnamefont {Henry}}, \ and\ \bibinfo {author} {\bibfnamefont
			{G.}~\bibnamefont {Lapertot}},\ }\href@noop {} {\bibfield  {journal}
		{\bibinfo  {journal} {Physica C}\ }\textbf {\bibinfo {volume} {185-189}},\
		\bibinfo {pages} {86} (\bibinfo {year} {1991})}\BibitemShut {NoStop}%
	\bibitem [{\citenamefont {Bourges}\ \emph {et~al.}(1996)\citenamefont
		{Bourges}, \citenamefont {Regnault}, \citenamefont {Sidis},\ and\
		\citenamefont {Vettier}}]{Bourges1996}%
	\BibitemOpen
	\bibfield  {author} {\bibinfo {author} {\bibfnamefont {P.}~\bibnamefont
			{Bourges}}, \bibinfo {author} {\bibfnamefont {L.~P.}\ \bibnamefont
			{Regnault}}, \bibinfo {author} {\bibfnamefont {Y.}~\bibnamefont {Sidis}}, \
		and\ \bibinfo {author} {\bibfnamefont {C.}~\bibnamefont {Vettier}},\
	}\href@noop {} {\bibfield  {journal} {\bibinfo  {journal} {Phys. Rev. B}\
		}\textbf {\bibinfo {volume} {53}},\ \bibinfo {pages} {876} (\bibinfo {year}
		{1996})},\ \bibinfo {note} {pRB}\BibitemShut {NoStop}%
	\bibitem [{\citenamefont {Dai}\ \emph {et~al.}(2000)\citenamefont {Dai},
		\citenamefont {Mook}, \citenamefont {Aeppli}, \citenamefont {Hayden},\ and\
		\citenamefont {Dogan}}]{Dai2000}%
	\BibitemOpen
	\bibfield  {author} {\bibinfo {author} {\bibfnamefont {P.}~\bibnamefont
			{Dai}}, \bibinfo {author} {\bibfnamefont {H.~A.}\ \bibnamefont {Mook}},
		\bibinfo {author} {\bibfnamefont {G.}~\bibnamefont {Aeppli}}, \bibinfo
		{author} {\bibfnamefont {S.~M.}\ \bibnamefont {Hayden}}, \ and\ \bibinfo
		{author} {\bibfnamefont {F.}~\bibnamefont {Dogan}},\ }\href@noop {}
	{\bibfield  {journal} {\bibinfo  {journal} {Nature}\ }\textbf {\bibinfo
			{volume} {406}},\ \bibinfo {pages} {965} (\bibinfo {year}
		{2000})}\BibitemShut {NoStop}%
	\bibitem [{\citenamefont {Fong}\ \emph {et~al.}(1999)\citenamefont {Fong},
		\citenamefont {Bourges}, \citenamefont {Sidis}, \citenamefont {Regnault},
		\citenamefont {Ivanov}, \citenamefont {Gu}, \citenamefont {Koshizuka},\ and\
		\citenamefont {Keimer}}]{Fong1999}%
	\BibitemOpen
	\bibfield  {author} {\bibinfo {author} {\bibfnamefont {H.~F.}\ \bibnamefont
			{Fong}}, \bibinfo {author} {\bibfnamefont {P.}~\bibnamefont {Bourges}},
		\bibinfo {author} {\bibfnamefont {Y.}~\bibnamefont {Sidis}}, \bibinfo
		{author} {\bibfnamefont {L.~P.}\ \bibnamefont {Regnault}}, \bibinfo {author}
		{\bibfnamefont {A.}~\bibnamefont {Ivanov}}, \bibinfo {author} {\bibfnamefont
			{G.~D.}\ \bibnamefont {Gu}}, \bibinfo {author} {\bibfnamefont
			{N.}~\bibnamefont {Koshizuka}}, \ and\ \bibinfo {author} {\bibfnamefont
			{B.}~\bibnamefont {Keimer}},\ }\href {\doibase 10.1038/19255} {\bibfield
		{journal} {\bibinfo  {journal} {Nature}\ }\textbf {\bibinfo {volume} {398}},\
		\bibinfo {pages} {588} (\bibinfo {year} {1999})}\BibitemShut {NoStop}%
	\bibitem [{\citenamefont {Hayden}\ \emph {et~al.}(2004)\citenamefont {Hayden},
		\citenamefont {Mook}, \citenamefont {Dai}, \citenamefont {Perring},\ and\
		\citenamefont {Dogan}}]{Hayden2004}%
	\BibitemOpen
	\bibfield  {author} {\bibinfo {author} {\bibfnamefont {S.~M.}\ \bibnamefont
			{Hayden}}, \bibinfo {author} {\bibfnamefont {H.~A.}\ \bibnamefont {Mook}},
		\bibinfo {author} {\bibfnamefont {P.}~\bibnamefont {Dai}}, \bibinfo {author}
		{\bibfnamefont {T.~G.}\ \bibnamefont {Perring}}, \ and\ \bibinfo {author}
		{\bibfnamefont {F.}~\bibnamefont {Dogan}},\ }\href {\doibase
		10.1038/nature02576} {\bibfield  {journal} {\bibinfo  {journal} {Nature}\
		}\textbf {\bibinfo {volume} {429}},\ \bibinfo {pages} {531} (\bibinfo {year}
		{2004})}\BibitemShut {NoStop}%
	\bibitem [{\citenamefont {Vignolle}\ \emph {et~al.}(2007)\citenamefont
		{Vignolle}, \citenamefont {Hayden}, \citenamefont {McMorrow}, \citenamefont
		{Ronnow}, \citenamefont {Lake}, \citenamefont {Frost},\ and\ \citenamefont
		{Perring}}]{Vignolle2007}%
	\BibitemOpen
	\bibfield  {author} {\bibinfo {author} {\bibfnamefont {B.}~\bibnamefont
			{Vignolle}}, \bibinfo {author} {\bibfnamefont {S.~M.}\ \bibnamefont
			{Hayden}}, \bibinfo {author} {\bibfnamefont {D.~F.}\ \bibnamefont
			{McMorrow}}, \bibinfo {author} {\bibfnamefont {H.~M.}\ \bibnamefont
			{Ronnow}}, \bibinfo {author} {\bibfnamefont {B.}~\bibnamefont {Lake}},
		\bibinfo {author} {\bibfnamefont {C.~D.}\ \bibnamefont {Frost}}, \ and\
		\bibinfo {author} {\bibfnamefont {T.~G.}\ \bibnamefont {Perring}},\ }\href
	{\doibase 10.1038/nphys546} {\bibfield  {journal} {\bibinfo  {journal} {Nat.
				Phys.}\ }\textbf {\bibinfo {volume} {3}},\ \bibinfo {pages} {163} (\bibinfo
		{year} {2007})}\BibitemShut {NoStop}%
	\bibitem [{\citenamefont {Hinkov}\ \emph {et~al.}(2007)\citenamefont {Hinkov},
		\citenamefont {Bourges}, \citenamefont {Pailhes}, \citenamefont {Sidis},
		\citenamefont {Ivanov}, \citenamefont {Frost}, \citenamefont {Perring},
		\citenamefont {Lin}, \citenamefont {Chen},\ and\ \citenamefont
		{Keimer}}]{Hinkov2007np}%
	\BibitemOpen
	\bibfield  {author} {\bibinfo {author} {\bibfnamefont {V.}~\bibnamefont
			{Hinkov}}, \bibinfo {author} {\bibfnamefont {P.}~\bibnamefont {Bourges}},
		\bibinfo {author} {\bibfnamefont {S.}~\bibnamefont {Pailhes}}, \bibinfo
		{author} {\bibfnamefont {Y.}~\bibnamefont {Sidis}}, \bibinfo {author}
		{\bibfnamefont {A.}~\bibnamefont {Ivanov}}, \bibinfo {author} {\bibfnamefont
			{C.~D.}\ \bibnamefont {Frost}}, \bibinfo {author} {\bibfnamefont {T.~G.}\
			\bibnamefont {Perring}}, \bibinfo {author} {\bibfnamefont {C.~T.}\
			\bibnamefont {Lin}}, \bibinfo {author} {\bibfnamefont {D.~P.}\ \bibnamefont
			{Chen}}, \ and\ \bibinfo {author} {\bibfnamefont {B.}~\bibnamefont
			{Keimer}},\ }\href {\doibase 10.1038/nphys720} {\bibfield  {journal}
		{\bibinfo  {journal} {Nat. Phys.}\ }\textbf {\bibinfo {volume} {3}},\
		\bibinfo {pages} {780} (\bibinfo {year} {2007})}\BibitemShut {NoStop}%
	\bibitem [{\citenamefont {Lumsden}\ and\ \citenamefont
		{Christianson}(2010)}]{lums10r}%
	\BibitemOpen
	\bibfield  {author} {\bibinfo {author} {\bibfnamefont {M.~D.}\ \bibnamefont
			{Lumsden}}\ and\ \bibinfo {author} {\bibfnamefont {A.~D.}\ \bibnamefont
			{Christianson}},\ }\href@noop {} {\bibfield  {journal} {\bibinfo  {journal}
			{J. Phys. Condens. Matter}\ }\textbf {\bibinfo {volume} {22}},\ \bibinfo
		{pages} {203203} (\bibinfo {year} {2010})}\BibitemShut {NoStop}%
	\bibitem [{\citenamefont {Christianson}\ \emph {et~al.}(2008)\citenamefont
		{Christianson}, \citenamefont {Goremychkin}, \citenamefont {Osborn},
		\citenamefont {Rosenkranz}, \citenamefont {Lumsden}, \citenamefont
		{Malliakas}, \citenamefont {Todorov}, \citenamefont {Claus}, \citenamefont
		{Chung}, \citenamefont {Kanatzidis}, \citenamefont {Bewley},\ and\
		\citenamefont {Guidi}}]{Christianson2008}%
	\BibitemOpen
	\bibfield  {author} {\bibinfo {author} {\bibfnamefont {A.~D.}\ \bibnamefont
			{Christianson}}, \bibinfo {author} {\bibfnamefont {E.~A.}\ \bibnamefont
			{Goremychkin}}, \bibinfo {author} {\bibfnamefont {R.}~\bibnamefont {Osborn}},
		\bibinfo {author} {\bibfnamefont {S.}~\bibnamefont {Rosenkranz}}, \bibinfo
		{author} {\bibfnamefont {M.~D.}\ \bibnamefont {Lumsden}}, \bibinfo {author}
		{\bibfnamefont {C.~D.}\ \bibnamefont {Malliakas}}, \bibinfo {author}
		{\bibfnamefont {I.~S.}\ \bibnamefont {Todorov}}, \bibinfo {author}
		{\bibfnamefont {H.}~\bibnamefont {Claus}}, \bibinfo {author} {\bibfnamefont
			{D.~Y.}\ \bibnamefont {Chung}}, \bibinfo {author} {\bibfnamefont {M.~G.}\
			\bibnamefont {Kanatzidis}}, \bibinfo {author} {\bibfnamefont {R.~I.}\
			\bibnamefont {Bewley}}, \ and\ \bibinfo {author} {\bibfnamefont
			{T.}~\bibnamefont {Guidi}},\ }\href {\doibase 10.1038/nature07625} {\bibfield
		{journal} {\bibinfo  {journal} {Nature}\ }\textbf {\bibinfo {volume}
			{456}},\ \bibinfo {pages} {930} (\bibinfo {year} {2008})}\BibitemShut
	{NoStop}%
	\bibitem [{\citenamefont {Christianson}\ \emph {et~al.}(2013)\citenamefont
		{Christianson}, \citenamefont {Lumsden}, \citenamefont {Marty}, \citenamefont
		{Wang}, \citenamefont {Calder}, \citenamefont {Abernathy}, \citenamefont
		{Stone}, \citenamefont {Mook}, \citenamefont {McGuire}, \citenamefont
		{Sefat}, \citenamefont {Sales}, \citenamefont {Mandrus},\ and\ \citenamefont
		{Goremychkin}}]{Christianson2013}%
	\BibitemOpen
	\bibfield  {author} {\bibinfo {author} {\bibfnamefont {A.~D.}\ \bibnamefont
			{Christianson}}, \bibinfo {author} {\bibfnamefont {M.~D.}\ \bibnamefont
			{Lumsden}}, \bibinfo {author} {\bibfnamefont {K.}~\bibnamefont {Marty}},
		\bibinfo {author} {\bibfnamefont {C.~H.}\ \bibnamefont {Wang}}, \bibinfo
		{author} {\bibfnamefont {S.}~\bibnamefont {Calder}}, \bibinfo {author}
		{\bibfnamefont {D.~L.}\ \bibnamefont {Abernathy}}, \bibinfo {author}
		{\bibfnamefont {M.~B.}\ \bibnamefont {Stone}}, \bibinfo {author}
		{\bibfnamefont {H.~A.}\ \bibnamefont {Mook}}, \bibinfo {author}
		{\bibfnamefont {M.~A.}\ \bibnamefont {McGuire}}, \bibinfo {author}
		{\bibfnamefont {A.~S.}\ \bibnamefont {Sefat}}, \bibinfo {author}
		{\bibfnamefont {B.~C.}\ \bibnamefont {Sales}}, \bibinfo {author}
		{\bibfnamefont {D.}~\bibnamefont {Mandrus}}, \ and\ \bibinfo {author}
		{\bibfnamefont {E.~A.}\ \bibnamefont {Goremychkin}},\ }\href@noop {}
	{\bibfield  {journal} {\bibinfo  {journal} {Phys. Rev. B}\ }\textbf {\bibinfo
			{volume} {87}},\ \bibinfo {pages} {224410} (\bibinfo {year}
		{2013})}\BibitemShut {NoStop}%
	\bibitem [{\citenamefont {Lumsden}\ \emph {et~al.}(2009)\citenamefont
		{Lumsden}, \citenamefont {Christianson}, \citenamefont {Parshall},
		\citenamefont {Stone}, \citenamefont {Nagler}, \citenamefont {MacDougall},
		\citenamefont {Mook}, \citenamefont {Lokshin}, \citenamefont {Egami},
		\citenamefont {Abernathy}, \citenamefont {Goremychkin}, \citenamefont
		{Osborn}, \citenamefont {McGuire}, \citenamefont {Sefat}, \citenamefont
		{Jin}, \citenamefont {Sales},\ and\ \citenamefont
		{Mandrus}}]{Lumsden2009prl}%
	\BibitemOpen
	\bibfield  {author} {\bibinfo {author} {\bibfnamefont {M.~D.}\ \bibnamefont
			{Lumsden}}, \bibinfo {author} {\bibfnamefont {A.~D.}\ \bibnamefont
			{Christianson}}, \bibinfo {author} {\bibfnamefont {D.}~\bibnamefont
			{Parshall}}, \bibinfo {author} {\bibfnamefont {M.~B.}\ \bibnamefont {Stone}},
		\bibinfo {author} {\bibfnamefont {S.~E.}\ \bibnamefont {Nagler}}, \bibinfo
		{author} {\bibfnamefont {G.~J.}\ \bibnamefont {MacDougall}}, \bibinfo
		{author} {\bibfnamefont {H.~A.}\ \bibnamefont {Mook}}, \bibinfo {author}
		{\bibfnamefont {K.}~\bibnamefont {Lokshin}}, \bibinfo {author} {\bibfnamefont
			{T.}~\bibnamefont {Egami}}, \bibinfo {author} {\bibfnamefont {D.~L.}\
			\bibnamefont {Abernathy}}, \bibinfo {author} {\bibfnamefont {E.~A.}\
			\bibnamefont {Goremychkin}}, \bibinfo {author} {\bibfnamefont
			{R.}~\bibnamefont {Osborn}}, \bibinfo {author} {\bibfnamefont {M.~A.}\
			\bibnamefont {McGuire}}, \bibinfo {author} {\bibfnamefont {A.~S.}\
			\bibnamefont {Sefat}}, \bibinfo {author} {\bibfnamefont {R.}~\bibnamefont
			{Jin}}, \bibinfo {author} {\bibfnamefont {B.~C.}\ \bibnamefont {Sales}}, \
		and\ \bibinfo {author} {\bibfnamefont {D.}~\bibnamefont {Mandrus}},\
	}\href@noop {} {\bibfield  {journal} {\bibinfo  {journal} {Phys. Rev. Lett.}\
		}\textbf {\bibinfo {volume} {102}},\ \bibinfo {pages} {107005} (\bibinfo
		{year} {2009})}\BibitemShut {NoStop}%
	\bibitem [{\citenamefont {Chi}\ \emph {et~al.}(2009)\citenamefont {Chi},
		\citenamefont {Schneidewind}, \citenamefont {Zhao}, \citenamefont {Harriger},
		\citenamefont {Li}, \citenamefont {Luo}, \citenamefont {Cao}, \citenamefont
		{Xu}, \citenamefont {Loewenhaupt}, \citenamefont {Hu},\ and\ \citenamefont
		{Dai}}]{Chis2009prl}%
	\BibitemOpen
	\bibfield  {author} {\bibinfo {author} {\bibfnamefont {S.}~\bibnamefont
			{Chi}}, \bibinfo {author} {\bibfnamefont {A.}~\bibnamefont {Schneidewind}},
		\bibinfo {author} {\bibfnamefont {J.}~\bibnamefont {Zhao}}, \bibinfo {author}
		{\bibfnamefont {L.~W.}\ \bibnamefont {Harriger}}, \bibinfo {author}
		{\bibfnamefont {L.}~\bibnamefont {Li}}, \bibinfo {author} {\bibfnamefont
			{Y.}~\bibnamefont {Luo}}, \bibinfo {author} {\bibfnamefont {G.}~\bibnamefont
			{Cao}}, \bibinfo {author} {\bibfnamefont {Z.~A.}\ \bibnamefont {Xu}},
		\bibinfo {author} {\bibfnamefont {M.}~\bibnamefont {Loewenhaupt}}, \bibinfo
		{author} {\bibfnamefont {J.}~\bibnamefont {Hu}}, \ and\ \bibinfo {author}
		{\bibfnamefont {P.}~\bibnamefont {Dai}},\ }\href@noop {} {\bibfield
		{journal} {\bibinfo  {journal} {Phys. Rev. Lett.}\ }\textbf {\bibinfo
			{volume} {102}},\ \bibinfo {pages} {107006} (\bibinfo {year}
		{2009})}\BibitemShut {NoStop}%
	\bibitem [{\citenamefont {Inosov}\ \emph {et~al.}(2010)\citenamefont {Inosov},
		\citenamefont {Park}, \citenamefont {Bourges}, \citenamefont {Sun},
		\citenamefont {Sidis}, \citenamefont {Schneidewind}, \citenamefont {Hradil},
		\citenamefont {Haug}, \citenamefont {Lin}, \citenamefont {Keimer},\ and\
		\citenamefont {Hinkov}}]{Inosov2010nf}%
	\BibitemOpen
	\bibfield  {author} {\bibinfo {author} {\bibfnamefont {D.~S.}\ \bibnamefont
			{Inosov}}, \bibinfo {author} {\bibfnamefont {J.~T.}\ \bibnamefont {Park}},
		\bibinfo {author} {\bibfnamefont {P.}~\bibnamefont {Bourges}}, \bibinfo
		{author} {\bibfnamefont {D.~L.}\ \bibnamefont {Sun}}, \bibinfo {author}
		{\bibfnamefont {Y.}~\bibnamefont {Sidis}}, \bibinfo {author} {\bibfnamefont
			{A.}~\bibnamefont {Schneidewind}}, \bibinfo {author} {\bibfnamefont
			{K.}~\bibnamefont {Hradil}}, \bibinfo {author} {\bibfnamefont
			{D.}~\bibnamefont {Haug}}, \bibinfo {author} {\bibfnamefont {C.~T.}\
			\bibnamefont {Lin}}, \bibinfo {author} {\bibfnamefont {B.}~\bibnamefont
			{Keimer}}, \ and\ \bibinfo {author} {\bibfnamefont {V.}~\bibnamefont
			{Hinkov}},\ }\href {\doibase 10.1038/nphys1483} {\bibfield  {journal}
		{\bibinfo  {journal} {Nat. Phys.}\ }\textbf {\bibinfo {volume} {6}},\
		\bibinfo {pages} {178} (\bibinfo {year} {2010})}\BibitemShut {NoStop}%
	\bibitem [{\citenamefont {Qiu}\ \emph {et~al.}(2009)\citenamefont {Qiu},
		\citenamefont {Bao}, \citenamefont {Zhao}, \citenamefont {Broholm},
		\citenamefont {Stanev}, \citenamefont {Tesanovic}, \citenamefont
		{Gasparovic}, \citenamefont {Chang}, \citenamefont {Hu}, \citenamefont
		{Qian}, \citenamefont {Fang},\ and\ \citenamefont {Mao}}]{Qiu2009}%
	\BibitemOpen
	\bibfield  {author} {\bibinfo {author} {\bibfnamefont {Y.}~\bibnamefont
			{Qiu}}, \bibinfo {author} {\bibfnamefont {W.}~\bibnamefont {Bao}}, \bibinfo
		{author} {\bibfnamefont {Y.}~\bibnamefont {Zhao}}, \bibinfo {author}
		{\bibfnamefont {C.}~\bibnamefont {Broholm}}, \bibinfo {author} {\bibfnamefont
			{V.}~\bibnamefont {Stanev}}, \bibinfo {author} {\bibfnamefont
			{Z.}~\bibnamefont {Tesanovic}}, \bibinfo {author} {\bibfnamefont {Y.~C.}\
			\bibnamefont {Gasparovic}}, \bibinfo {author} {\bibfnamefont
			{S.}~\bibnamefont {Chang}}, \bibinfo {author} {\bibfnamefont
			{J.}~\bibnamefont {Hu}}, \bibinfo {author} {\bibfnamefont {B.}~\bibnamefont
			{Qian}}, \bibinfo {author} {\bibfnamefont {M.}~\bibnamefont {Fang}}, \ and\
		\bibinfo {author} {\bibfnamefont {Z.}~\bibnamefont {Mao}},\ }\href@noop {}
	{\bibfield  {journal} {\bibinfo  {journal} {Phys. Rev. Lett.}\ }\textbf
		{\bibinfo {volume} {103}},\ \bibinfo {pages} {067008} (\bibinfo {year}
		{2009})}\BibitemShut {NoStop}%
	\bibitem [{\citenamefont {Wen}\ \emph {et~al.}(2010)\citenamefont {Wen},
		\citenamefont {Xu}, \citenamefont {Xu}, \citenamefont {Lin}, \citenamefont
		{Li}, \citenamefont {Chen}, \citenamefont {Chi}, \citenamefont {Gu},\ and\
		\citenamefont {Tranquada}}]{Wen2010H}%
	\BibitemOpen
	\bibfield  {author} {\bibinfo {author} {\bibfnamefont {J.}~\bibnamefont
			{Wen}}, \bibinfo {author} {\bibfnamefont {G.}~\bibnamefont {Xu}}, \bibinfo
		{author} {\bibfnamefont {Z.}~\bibnamefont {Xu}}, \bibinfo {author}
		{\bibfnamefont {Z.~W.}\ \bibnamefont {Lin}}, \bibinfo {author} {\bibfnamefont
			{Q.}~\bibnamefont {Li}}, \bibinfo {author} {\bibfnamefont {Y.}~\bibnamefont
			{Chen}}, \bibinfo {author} {\bibfnamefont {S.}~\bibnamefont {Chi}}, \bibinfo
		{author} {\bibfnamefont {G.}~\bibnamefont {Gu}}, \ and\ \bibinfo {author}
		{\bibfnamefont {J.~M.}\ \bibnamefont {Tranquada}},\ }\href@noop {} {\bibfield
		{journal} {\bibinfo  {journal} {Phys. Rev. B}\ }\textbf {\bibinfo {volume}
			{81}},\ \bibinfo {pages} {100513(R)} (\bibinfo {year} {2010})}\BibitemShut
	{NoStop}%
	\bibitem [{\citenamefont {Yu}\ \emph {et~al.}(2009)\citenamefont {Yu},
		\citenamefont {Li}, \citenamefont {Motoyama},\ and\ \citenamefont
		{Greven}}]{yu09}%
	\BibitemOpen
	\bibfield  {author} {\bibinfo {author} {\bibfnamefont {G.}~\bibnamefont
			{Yu}}, \bibinfo {author} {\bibfnamefont {Y.}~\bibnamefont {Li}}, \bibinfo
		{author} {\bibfnamefont {E.~M.}\ \bibnamefont {Motoyama}}, \ and\ \bibinfo
		{author} {\bibfnamefont {M.}~\bibnamefont {Greven}},\ }\href@noop {}
	{\bibfield  {journal} {\bibinfo  {journal} {Nat. Phys.}\ }\textbf {\bibinfo
			{volume} {5}},\ \bibinfo {pages} {873} (\bibinfo {year} {2009})}\BibitemShut
	{NoStop}%
	\bibitem [{\citenamefont {Li}\ \emph {et~al.}(2009)\citenamefont {Li},
		\citenamefont {Chen}, \citenamefont {Chang}, \citenamefont {Lynn},
		\citenamefont {Li}, \citenamefont {Luo}, \citenamefont {Cao}, \citenamefont
		{Xu},\ and\ \citenamefont {Dai}}]{Lis2009prb}%
	\BibitemOpen
	\bibfield  {author} {\bibinfo {author} {\bibfnamefont {S.}~\bibnamefont
			{Li}}, \bibinfo {author} {\bibfnamefont {Y.}~\bibnamefont {Chen}}, \bibinfo
		{author} {\bibfnamefont {S.}~\bibnamefont {Chang}}, \bibinfo {author}
		{\bibfnamefont {J.~W.}\ \bibnamefont {Lynn}}, \bibinfo {author}
		{\bibfnamefont {L.}~\bibnamefont {Li}}, \bibinfo {author} {\bibfnamefont
			{Y.}~\bibnamefont {Luo}}, \bibinfo {author} {\bibfnamefont {G.}~\bibnamefont
			{Cao}}, \bibinfo {author} {\bibfnamefont {Z.}~\bibnamefont {Xu}}, \ and\
		\bibinfo {author} {\bibfnamefont {P.}~\bibnamefont {Dai}},\ }\href@noop {}
	{\bibfield  {journal} {\bibinfo  {journal} {Phys. Rev. B}\ }\textbf {\bibinfo
			{volume} {79}},\ \bibinfo {pages} {174527} (\bibinfo {year}
		{2009})}\BibitemShut {NoStop}%
	\bibitem [{\citenamefont {Wakimoto}\ \emph {et~al.}(2010)\citenamefont
		{Wakimoto}, \citenamefont {Kodama}, \citenamefont {Ishikado}, \citenamefont
		{Matsuda}, \citenamefont {Kajimoto}, \citenamefont {Arai}, \citenamefont
		{Kakurai}, \citenamefont {Esaka}, \citenamefont {Iyo}, \citenamefont {Kito},
		\citenamefont {Eisaki},\ and\ \citenamefont {Shamoto}}]{Wakimoto2010}%
	\BibitemOpen
	\bibfield  {author} {\bibinfo {author} {\bibfnamefont {S.}~\bibnamefont
			{Wakimoto}}, \bibinfo {author} {\bibfnamefont {K.}~\bibnamefont {Kodama}},
		\bibinfo {author} {\bibfnamefont {M.}~\bibnamefont {Ishikado}}, \bibinfo
		{author} {\bibfnamefont {M.}~\bibnamefont {Matsuda}}, \bibinfo {author}
		{\bibfnamefont {R.}~\bibnamefont {Kajimoto}}, \bibinfo {author}
		{\bibfnamefont {M.}~\bibnamefont {Arai}}, \bibinfo {author} {\bibfnamefont
			{K.}~\bibnamefont {Kakurai}}, \bibinfo {author} {\bibfnamefont
			{F.}~\bibnamefont {Esaka}}, \bibinfo {author} {\bibfnamefont
			{A.}~\bibnamefont {Iyo}}, \bibinfo {author} {\bibfnamefont {H.}~\bibnamefont
			{Kito}}, \bibinfo {author} {\bibfnamefont {H.}~\bibnamefont {Eisaki}}, \ and\
		\bibinfo {author} {\bibfnamefont {S.-i.}\ \bibnamefont {Shamoto}},\ }\href
	{\doibase 10.1143/JPSJ.79.074715} {\bibfield  {journal} {\bibinfo  {journal}
			{J. Phys. Soc. Jpn.}\ }\textbf {\bibinfo {volume} {79}},\ \bibinfo {pages}
		{074715} (\bibinfo {year} {2010})}\BibitemShut {NoStop}%
	\bibitem [{\citenamefont {Zhang}\ \emph {et~al.}(2013)\citenamefont {Zhang},
		\citenamefont {Yu}, \citenamefont {Su}, \citenamefont {Song}, \citenamefont
		{Wang}, \citenamefont {Tan}, \citenamefont {Egami}, \citenamefont
		{Fernandez-Baca}, \citenamefont {Faulhaber}, \citenamefont {Si},\ and\
		\citenamefont {Dai}}]{Zhang2013}%
	\BibitemOpen
	\bibfield  {author} {\bibinfo {author} {\bibfnamefont {C.}~\bibnamefont
			{Zhang}}, \bibinfo {author} {\bibfnamefont {R.}~\bibnamefont {Yu}}, \bibinfo
		{author} {\bibfnamefont {Y.}~\bibnamefont {Su}}, \bibinfo {author}
		{\bibfnamefont {Y.}~\bibnamefont {Song}}, \bibinfo {author} {\bibfnamefont
			{M.}~\bibnamefont {Wang}}, \bibinfo {author} {\bibfnamefont {G.}~\bibnamefont
			{Tan}}, \bibinfo {author} {\bibfnamefont {T.}~\bibnamefont {Egami}}, \bibinfo
		{author} {\bibfnamefont {J.~A.}\ \bibnamefont {Fernandez-Baca}}, \bibinfo
		{author} {\bibfnamefont {E.}~\bibnamefont {Faulhaber}}, \bibinfo {author}
		{\bibfnamefont {Q.}~\bibnamefont {Si}}, \ and\ \bibinfo {author}
		{\bibfnamefont {P.}~\bibnamefont {Dai}},\ }\href@noop {} {\bibfield
		{journal} {\bibinfo  {journal} {Phys. Rev. Lett.}\ }\textbf {\bibinfo
			{volume} {111}},\ \bibinfo {pages} {207002} (\bibinfo {year}
		{2013})}\BibitemShut {NoStop}%
	\bibitem [{\citenamefont {Zhang}\ \emph {et~al.}(2015)\citenamefont {Zhang},
		\citenamefont {Park}, \citenamefont {Lu}, \citenamefont {Yu}, \citenamefont
		{Li}, \citenamefont {Zhang}, \citenamefont {Zhao}, \citenamefont {Lynn},
		\citenamefont {Si},\ and\ \citenamefont {Dai}}]{Zhang2015}%
	\BibitemOpen
	\bibfield  {author} {\bibinfo {author} {\bibfnamefont {C.}~\bibnamefont
			{Zhang}}, \bibinfo {author} {\bibfnamefont {J.~T.}\ \bibnamefont {Park}},
		\bibinfo {author} {\bibfnamefont {X.}~\bibnamefont {Lu}}, \bibinfo {author}
		{\bibfnamefont {R.}~\bibnamefont {Yu}}, \bibinfo {author} {\bibfnamefont
			{Y.}~\bibnamefont {Li}}, \bibinfo {author} {\bibfnamefont {W.}~\bibnamefont
			{Zhang}}, \bibinfo {author} {\bibfnamefont {Y.}~\bibnamefont {Zhao}},
		\bibinfo {author} {\bibfnamefont {J.~W.}\ \bibnamefont {Lynn}}, \bibinfo
		{author} {\bibfnamefont {Q.}~\bibnamefont {Si}}, \ and\ \bibinfo {author}
		{\bibfnamefont {P.}~\bibnamefont {Dai}},\ }\href@noop {} {\bibfield
		{journal} {\bibinfo  {journal} {Phys. Rev. B}\ }\textbf {\bibinfo {volume}
			{91}},\ \bibinfo {pages} {104520} (\bibinfo {year} {2015})}\BibitemShut
	{NoStop}%
	\bibitem [{\citenamefont {Dai}(2015)}]{Dai2015}%
	\BibitemOpen
	\bibfield  {author} {\bibinfo {author} {\bibfnamefont {P.}~\bibnamefont
			{Dai}},\ }\href@noop {} {\bibfield  {journal} {\bibinfo  {journal} {Rev. Mod.
				Phys.}\ }\textbf {\bibinfo {volume} {87}},\ \bibinfo {pages} {855} (\bibinfo
		{year} {2015})}\BibitemShut {NoStop}%
	\bibitem [{\citenamefont {Bao}\ \emph {et~al.}(2009)\citenamefont {Bao},
		\citenamefont {Qiu}, \citenamefont {Huang}, \citenamefont {Green},
		\citenamefont {Zajdel}, \citenamefont {Fitzsimmons}, \citenamefont
		{Zhernenkov}, \citenamefont {Chang}, \citenamefont {Fang}, \citenamefont
		{Qian}, \citenamefont {Vehstedt}, \citenamefont {Yang}, \citenamefont {Pham},
		\citenamefont {Spinu},\ and\ \citenamefont {Mao}}]{Bao2009}%
	\BibitemOpen
	\bibfield  {author} {\bibinfo {author} {\bibfnamefont {W.}~\bibnamefont
			{Bao}}, \bibinfo {author} {\bibfnamefont {Y.}~\bibnamefont {Qiu}}, \bibinfo
		{author} {\bibfnamefont {Q.}~\bibnamefont {Huang}}, \bibinfo {author}
		{\bibfnamefont {M.~A.}\ \bibnamefont {Green}}, \bibinfo {author}
		{\bibfnamefont {P.}~\bibnamefont {Zajdel}}, \bibinfo {author} {\bibfnamefont
			{M.~R.}\ \bibnamefont {Fitzsimmons}}, \bibinfo {author} {\bibfnamefont
			{M.}~\bibnamefont {Zhernenkov}}, \bibinfo {author} {\bibfnamefont
			{S.}~\bibnamefont {Chang}}, \bibinfo {author} {\bibfnamefont
			{M.}~\bibnamefont {Fang}}, \bibinfo {author} {\bibfnamefont {B.}~\bibnamefont
			{Qian}}, \bibinfo {author} {\bibfnamefont {E.~K.}\ \bibnamefont {Vehstedt}},
		\bibinfo {author} {\bibfnamefont {J.}~\bibnamefont {Yang}}, \bibinfo {author}
		{\bibfnamefont {H.~M.}\ \bibnamefont {Pham}}, \bibinfo {author}
		{\bibfnamefont {L.}~\bibnamefont {Spinu}}, \ and\ \bibinfo {author}
		{\bibfnamefont {Z.~Q.}\ \bibnamefont {Mao}},\ }\href@noop {} {\bibfield
		{journal} {\bibinfo  {journal} {Phys. Rev. Lett.}\ }\textbf {\bibinfo
			{volume} {102}},\ \bibinfo {pages} {247001} (\bibinfo {year}
		{2009})}\BibitemShut {NoStop}%
	\bibitem [{\citenamefont {Lumsden}\ \emph {et~al.}(2010)\citenamefont
		{Lumsden}, \citenamefont {Christianson}, \citenamefont {Goremychkin},
		\citenamefont {Nagler}, \citenamefont {Mook}, \citenamefont {Stone},
		\citenamefont {Abernathy}, \citenamefont {Guidi}, \citenamefont {MacDougall},
		\citenamefont {de~la Cruz}, \citenamefont {Sefat}, \citenamefont {McGuire},
		\citenamefont {Sales},\ and\ \citenamefont {Mandrus}}]{Lumsden2010nf}%
	\BibitemOpen
	\bibfield  {author} {\bibinfo {author} {\bibfnamefont {M.~D.}\ \bibnamefont
			{Lumsden}}, \bibinfo {author} {\bibfnamefont {A.~D.}\ \bibnamefont
			{Christianson}}, \bibinfo {author} {\bibfnamefont {E.~A.}\ \bibnamefont
			{Goremychkin}}, \bibinfo {author} {\bibfnamefont {S.~E.}\ \bibnamefont
			{Nagler}}, \bibinfo {author} {\bibfnamefont {H.~A.}\ \bibnamefont {Mook}},
		\bibinfo {author} {\bibfnamefont {M.~B.}\ \bibnamefont {Stone}}, \bibinfo
		{author} {\bibfnamefont {D.~L.}\ \bibnamefont {Abernathy}}, \bibinfo {author}
		{\bibfnamefont {T.}~\bibnamefont {Guidi}}, \bibinfo {author} {\bibfnamefont
			{G.~J.}\ \bibnamefont {MacDougall}}, \bibinfo {author} {\bibfnamefont
			{C.}~\bibnamefont {de~la Cruz}}, \bibinfo {author} {\bibfnamefont {A.~S.}\
			\bibnamefont {Sefat}}, \bibinfo {author} {\bibfnamefont {M.~A.}\ \bibnamefont
			{McGuire}}, \bibinfo {author} {\bibfnamefont {B.~C.}\ \bibnamefont {Sales}},
		\ and\ \bibinfo {author} {\bibfnamefont {D.}~\bibnamefont {Mandrus}},\ }\href
	{\doibase 10.1038/nphys1512} {\bibfield  {journal} {\bibinfo  {journal} {Nat.
				Phys.}\ }\textbf {\bibinfo {volume} {6}},\ \bibinfo {pages} {182} (\bibinfo
		{year} {2010})}\BibitemShut {NoStop}%
	\bibitem [{\citenamefont {Xu}\ \emph {et~al.}(2010)\citenamefont {Xu},
		\citenamefont {Wen}, \citenamefont {Xu}, \citenamefont {Jie}, \citenamefont
		{Lin}, \citenamefont {Li}, \citenamefont {Chi}, \citenamefont {Singh},
		\citenamefont {Gu},\ and\ \citenamefont {Tranquada}}]{zxu2010fetese1}%
	\BibitemOpen
	\bibfield  {author} {\bibinfo {author} {\bibfnamefont {Z.}~\bibnamefont
			{Xu}}, \bibinfo {author} {\bibfnamefont {J.}~\bibnamefont {Wen}}, \bibinfo
		{author} {\bibfnamefont {G.}~\bibnamefont {Xu}}, \bibinfo {author}
		{\bibfnamefont {Q.}~\bibnamefont {Jie}}, \bibinfo {author} {\bibfnamefont
			{Z.}~\bibnamefont {Lin}}, \bibinfo {author} {\bibfnamefont {Q.}~\bibnamefont
			{Li}}, \bibinfo {author} {\bibfnamefont {S.}~\bibnamefont {Chi}}, \bibinfo
		{author} {\bibfnamefont {D.~K.}\ \bibnamefont {Singh}}, \bibinfo {author}
		{\bibfnamefont {G.}~\bibnamefont {Gu}}, \ and\ \bibinfo {author}
		{\bibfnamefont {J.~M.}\ \bibnamefont {Tranquada}},\ }\href@noop {} {\bibfield
		{journal} {\bibinfo  {journal} {Phys. Rev. B}\ }\textbf {\bibinfo {volume}
			{82}},\ \bibinfo {pages} {104525} (\bibinfo {year} {2010})}\BibitemShut
	{NoStop}%
	\bibitem [{\citenamefont {Xu}\ \emph {et~al.}(2016)\citenamefont {Xu},
		\citenamefont {Schneeloch}, \citenamefont {Wen}, \citenamefont {Bozin},
		\citenamefont {Granroth}, \citenamefont {Winn}, \citenamefont {Feygenson},
		\citenamefont {Birgeneau}, \citenamefont {Gu}, \citenamefont {Zaliznyak},
		\citenamefont {Tranquada},\ and\ \citenamefont {Xu}}]{Xu2016}%
	\BibitemOpen
	\bibfield  {author} {\bibinfo {author} {\bibfnamefont {Z.}~\bibnamefont
			{Xu}}, \bibinfo {author} {\bibfnamefont {J.~A.}\ \bibnamefont {Schneeloch}},
		\bibinfo {author} {\bibfnamefont {J.}~\bibnamefont {Wen}}, \bibinfo {author}
		{\bibfnamefont {E.~S.}\ \bibnamefont {Bozin}}, \bibinfo {author}
		{\bibfnamefont {G.~E.}\ \bibnamefont {Granroth}}, \bibinfo {author}
		{\bibfnamefont {B.~L.}\ \bibnamefont {Winn}}, \bibinfo {author}
		{\bibfnamefont {M.}~\bibnamefont {Feygenson}}, \bibinfo {author}
		{\bibfnamefont {R.~J.}\ \bibnamefont {Birgeneau}}, \bibinfo {author}
		{\bibfnamefont {G.}~\bibnamefont {Gu}}, \bibinfo {author} {\bibfnamefont
			{I.~A.}\ \bibnamefont {Zaliznyak}}, \bibinfo {author} {\bibfnamefont {J.~M.}\
			\bibnamefont {Tranquada}}, \ and\ \bibinfo {author} {\bibfnamefont
			{G.}~\bibnamefont {Xu}},\ }\href@noop {} {\bibfield  {journal} {\bibinfo
			{journal} {Phys. Rev. B}\ }\textbf {\bibinfo {volume} {93}},\ \bibinfo
		{pages} {104517} (\bibinfo {year} {2016})}\BibitemShut {NoStop}%
	\bibitem [{\citenamefont {Lipscombe}\ \emph {et~al.}(2011)\citenamefont
		{Lipscombe}, \citenamefont {Chen}, \citenamefont {Fang}, \citenamefont
		{Perring}, \citenamefont {Abernathy}, \citenamefont {Christianson},
		\citenamefont {Egami}, \citenamefont {Wang}, \citenamefont {Hu},\ and\
		\citenamefont {Dai}}]{Lipscombe2011}%
	\BibitemOpen
	\bibfield  {author} {\bibinfo {author} {\bibfnamefont {O.~J.}\ \bibnamefont
			{Lipscombe}}, \bibinfo {author} {\bibfnamefont {G.~F.}\ \bibnamefont {Chen}},
		\bibinfo {author} {\bibfnamefont {C.}~\bibnamefont {Fang}}, \bibinfo {author}
		{\bibfnamefont {T.~G.}\ \bibnamefont {Perring}}, \bibinfo {author}
		{\bibfnamefont {D.~L.}\ \bibnamefont {Abernathy}}, \bibinfo {author}
		{\bibfnamefont {A.~D.}\ \bibnamefont {Christianson}}, \bibinfo {author}
		{\bibfnamefont {T.}~\bibnamefont {Egami}}, \bibinfo {author} {\bibfnamefont
			{N.}~\bibnamefont {Wang}}, \bibinfo {author} {\bibfnamefont {J.}~\bibnamefont
			{Hu}}, \ and\ \bibinfo {author} {\bibfnamefont {P.}~\bibnamefont {Dai}},\
	}\href@noop {} {\bibfield  {journal} {\bibinfo  {journal} {Phys. Rev. Lett.}\
		}\textbf {\bibinfo {volume} {106}},\ \bibinfo {pages} {057004} (\bibinfo
		{year} {2011})}\BibitemShut {NoStop}%
	\bibitem [{\citenamefont {Zaliznyak}\ \emph {et~al.}(2011)\citenamefont
		{Zaliznyak}, \citenamefont {Xu}, \citenamefont {Tranquada}, \citenamefont
		{Gu}, \citenamefont {Tsvelik},\ and\ \citenamefont {Stone}}]{Zaliznyak2011}%
	\BibitemOpen
	\bibfield  {author} {\bibinfo {author} {\bibfnamefont {I.~A.}\ \bibnamefont
			{Zaliznyak}}, \bibinfo {author} {\bibfnamefont {Z.}~\bibnamefont {Xu}},
		\bibinfo {author} {\bibfnamefont {J.~M.}\ \bibnamefont {Tranquada}}, \bibinfo
		{author} {\bibfnamefont {G.}~\bibnamefont {Gu}}, \bibinfo {author}
		{\bibfnamefont {A.~M.}\ \bibnamefont {Tsvelik}}, \ and\ \bibinfo {author}
		{\bibfnamefont {M.~B.}\ \bibnamefont {Stone}},\ }\href@noop {} {\bibfield
		{journal} {\bibinfo  {journal} {Phys. Rev. Lett.}\ }\textbf {\bibinfo
			{volume} {107}},\ \bibinfo {pages} {216403} (\bibinfo {year}
		{2011})}\BibitemShut {NoStop}%
	\bibitem [{\citenamefont {Lee}\ \emph {et~al.}(2010)\citenamefont {Lee},
		\citenamefont {Xu}, \citenamefont {Ku}, \citenamefont {Wen}, \citenamefont
		{Lee}, \citenamefont {Katayama}, \citenamefont {Xu}, \citenamefont {Ji},
		\citenamefont {Lin}, \citenamefont {Gu}, \citenamefont {Yang}, \citenamefont
		{Johnson}, \citenamefont {Pan}, \citenamefont {Valla}, \citenamefont
		{Fujita}, \citenamefont {Sato}, \citenamefont {Chang}, \citenamefont
		{Yamada},\ and\ \citenamefont {Tranquada}}]{Lee2010}%
	\BibitemOpen
	\bibfield  {author} {\bibinfo {author} {\bibfnamefont {S.~H.}\ \bibnamefont
			{Lee}}, \bibinfo {author} {\bibfnamefont {G.}~\bibnamefont {Xu}}, \bibinfo
		{author} {\bibfnamefont {W.}~\bibnamefont {Ku}}, \bibinfo {author}
		{\bibfnamefont {J.~S.}\ \bibnamefont {Wen}}, \bibinfo {author} {\bibfnamefont
			{C.~C.}\ \bibnamefont {Lee}}, \bibinfo {author} {\bibfnamefont
			{N.}~\bibnamefont {Katayama}}, \bibinfo {author} {\bibfnamefont {Z.~J.}\
			\bibnamefont {Xu}}, \bibinfo {author} {\bibfnamefont {S.}~\bibnamefont {Ji}},
		\bibinfo {author} {\bibfnamefont {Z.~W.}\ \bibnamefont {Lin}}, \bibinfo
		{author} {\bibfnamefont {G.~D.}\ \bibnamefont {Gu}}, \bibinfo {author}
		{\bibfnamefont {H.~B.}\ \bibnamefont {Yang}}, \bibinfo {author}
		{\bibfnamefont {P.~D.}\ \bibnamefont {Johnson}}, \bibinfo {author}
		{\bibfnamefont {Z.~H.}\ \bibnamefont {Pan}}, \bibinfo {author} {\bibfnamefont
			{T.}~\bibnamefont {Valla}}, \bibinfo {author} {\bibfnamefont
			{M.}~\bibnamefont {Fujita}}, \bibinfo {author} {\bibfnamefont {T.~J.}\
			\bibnamefont {Sato}}, \bibinfo {author} {\bibfnamefont {S.}~\bibnamefont
			{Chang}}, \bibinfo {author} {\bibfnamefont {K.}~\bibnamefont {Yamada}}, \
		and\ \bibinfo {author} {\bibfnamefont {J.~M.}\ \bibnamefont {Tranquada}},\
	}\href@noop {} {\bibfield  {journal} {\bibinfo  {journal} {Phys. Rev. B}\
		}\textbf {\bibinfo {volume} {81}},\ \bibinfo {pages} {220502(R)} (\bibinfo
		{year} {2010})}\BibitemShut {NoStop}%
	\bibitem [{\citenamefont {Xu}\ \emph {et~al.}(2017)\citenamefont {Xu},
		\citenamefont {Schneeloch}, \citenamefont {Wen}, \citenamefont {Winn},
		\citenamefont {Granroth}, \citenamefont {Zhao}, \citenamefont {Gu},
		\citenamefont {Zaliznyak}, \citenamefont {Tranquada}, \citenamefont
		{Birgeneau},\ and\ \citenamefont {Xu}}]{Xu2017}%
	\BibitemOpen
	\bibfield  {author} {\bibinfo {author} {\bibfnamefont {Z.}~\bibnamefont
			{Xu}}, \bibinfo {author} {\bibfnamefont {J.~A.}\ \bibnamefont {Schneeloch}},
		\bibinfo {author} {\bibfnamefont {J.}~\bibnamefont {Wen}}, \bibinfo {author}
		{\bibfnamefont {B.~L.}\ \bibnamefont {Winn}}, \bibinfo {author}
		{\bibfnamefont {G.~E.}\ \bibnamefont {Granroth}}, \bibinfo {author}
		{\bibfnamefont {Y.}~\bibnamefont {Zhao}}, \bibinfo {author} {\bibfnamefont
			{G.}~\bibnamefont {Gu}}, \bibinfo {author} {\bibfnamefont {I.}~\bibnamefont
			{Zaliznyak}}, \bibinfo {author} {\bibfnamefont {J.~M.}\ \bibnamefont
			{Tranquada}}, \bibinfo {author} {\bibfnamefont {R.~J.}\ \bibnamefont
			{Birgeneau}}, \ and\ \bibinfo {author} {\bibfnamefont {G.}~\bibnamefont
			{Xu}},\ }\href@noop {} {\bibfield  {journal} {\bibinfo  {journal} {Phys. Rev.
				B}\ }\textbf {\bibinfo {volume} {96}},\ \bibinfo {pages} {134505} (\bibinfo
		{year} {2017})}\BibitemShut {NoStop}%
	\bibitem [{\citenamefont {Martinelli}\ \emph {et~al.}(2010)\citenamefont
		{Martinelli}, \citenamefont {Palenzona}, \citenamefont {Tropeano},
		\citenamefont {Ferdeghini}, \citenamefont {Putti}, \citenamefont {Cimberle},
		\citenamefont {Nguyen}, \citenamefont {Affronte},\ and\ \citenamefont
		{Ritter}}]{mart10b}%
	\BibitemOpen
	\bibfield  {author} {\bibinfo {author} {\bibfnamefont {A.}~\bibnamefont
			{Martinelli}}, \bibinfo {author} {\bibfnamefont {A.}~\bibnamefont
			{Palenzona}}, \bibinfo {author} {\bibfnamefont {M.}~\bibnamefont {Tropeano}},
		\bibinfo {author} {\bibfnamefont {C.}~\bibnamefont {Ferdeghini}}, \bibinfo
		{author} {\bibfnamefont {M.}~\bibnamefont {Putti}}, \bibinfo {author}
		{\bibfnamefont {M.~R.}\ \bibnamefont {Cimberle}}, \bibinfo {author}
		{\bibfnamefont {T.~D.}\ \bibnamefont {Nguyen}}, \bibinfo {author}
		{\bibfnamefont {M.}~\bibnamefont {Affronte}}, \ and\ \bibinfo {author}
		{\bibfnamefont {C.}~\bibnamefont {Ritter}},\ }\href@noop {} {\bibfield
		{journal} {\bibinfo  {journal} {Phys. Rev. B}\ }\textbf {\bibinfo {volume}
			{81}},\ \bibinfo {pages} {094115} (\bibinfo {year} {2010})}\BibitemShut
	{NoStop}%
	\bibitem [{\citenamefont {Katayama}\ \emph {et~al.}(2010)\citenamefont
		{Katayama}, \citenamefont {Ji}, \citenamefont {Louca}, \citenamefont {Lee},
		\citenamefont {Fujita}, \citenamefont {Sato}, \citenamefont {Wen},
		\citenamefont {Xu}, \citenamefont {Gu}, \citenamefont {Xu}, \citenamefont
		{Lin}, \citenamefont {Enoki}, \citenamefont {Chang}, \citenamefont {Yamada},\
		and\ \citenamefont {Tranquada}}]{Katayama2010}%
	\BibitemOpen
	\bibfield  {author} {\bibinfo {author} {\bibfnamefont {N.}~\bibnamefont
			{Katayama}}, \bibinfo {author} {\bibfnamefont {S.}~\bibnamefont {Ji}},
		\bibinfo {author} {\bibfnamefont {D.}~\bibnamefont {Louca}}, \bibinfo
		{author} {\bibfnamefont {S.-H.}\ \bibnamefont {Lee}}, \bibinfo {author}
		{\bibfnamefont {M.}~\bibnamefont {Fujita}}, \bibinfo {author} {\bibfnamefont
			{T.~J.}\ \bibnamefont {Sato}}, \bibinfo {author} {\bibfnamefont {J.~S.}\
			\bibnamefont {Wen}}, \bibinfo {author} {\bibfnamefont {Z.~J.}\ \bibnamefont
			{Xu}}, \bibinfo {author} {\bibfnamefont {G.~D.}\ \bibnamefont {Gu}}, \bibinfo
		{author} {\bibfnamefont {G.}~\bibnamefont {Xu}}, \bibinfo {author}
		{\bibfnamefont {Z.~W.}\ \bibnamefont {Lin}}, \bibinfo {author} {\bibfnamefont
			{M.}~\bibnamefont {Enoki}}, \bibinfo {author} {\bibfnamefont
			{S.}~\bibnamefont {Chang}}, \bibinfo {author} {\bibfnamefont
			{K.}~\bibnamefont {Yamada}}, \ and\ \bibinfo {author} {\bibfnamefont {J.~M.}\
			\bibnamefont {Tranquada}},\ }\href@noop {} {\bibfield  {journal} {\bibinfo
			{journal} {J. Phys. Soc. Jpn.}\ }\textbf {\bibinfo {volume} {79}},\ \bibinfo
		{pages} {113702} (\bibinfo {year} {2010})}\BibitemShut {NoStop}%
	\bibitem [{\citenamefont {Liu}\ \emph {et~al.}(2010)\citenamefont {Liu},
		\citenamefont {Hu}, \citenamefont {Qian}, \citenamefont {Fobes},
		\citenamefont {Mao}, \citenamefont {Bao}, \citenamefont {Reehuis},
		\citenamefont {Kimber}, \citenamefont {Proke拧}, \citenamefont {Matas},
		\citenamefont {Argyriou}, \citenamefont {Hiess}, \citenamefont {Rotaru},
		\citenamefont {Pham}, \citenamefont {Spinu}, \citenamefont {Qiu},
		\citenamefont {Thampy}, \citenamefont {Savici}, \citenamefont {Rodriguez},\
		and\ \citenamefont {Broholm}}]{liu10}%
	\BibitemOpen
	\bibfield  {author} {\bibinfo {author} {\bibfnamefont {T.~J.}\ \bibnamefont
			{Liu}}, \bibinfo {author} {\bibfnamefont {J.}~\bibnamefont {Hu}}, \bibinfo
		{author} {\bibfnamefont {B.}~\bibnamefont {Qian}}, \bibinfo {author}
		{\bibfnamefont {D.}~\bibnamefont {Fobes}}, \bibinfo {author} {\bibfnamefont
			{Z.~Q.}\ \bibnamefont {Mao}}, \bibinfo {author} {\bibfnamefont
			{W.}~\bibnamefont {Bao}}, \bibinfo {author} {\bibfnamefont {M.}~\bibnamefont
			{Reehuis}}, \bibinfo {author} {\bibfnamefont {S.~A.~J.}\ \bibnamefont
			{Kimber}}, \bibinfo {author} {\bibfnamefont {K.}~\bibnamefont {Proke拧}},
		\bibinfo {author} {\bibfnamefont {S.}~\bibnamefont {Matas}}, \bibinfo
		{author} {\bibfnamefont {D.~N.}\ \bibnamefont {Argyriou}}, \bibinfo {author}
		{\bibfnamefont {A.}~\bibnamefont {Hiess}}, \bibinfo {author} {\bibfnamefont
			{A.}~\bibnamefont {Rotaru}}, \bibinfo {author} {\bibfnamefont
			{H.}~\bibnamefont {Pham}}, \bibinfo {author} {\bibfnamefont {L.}~\bibnamefont
			{Spinu}}, \bibinfo {author} {\bibfnamefont {Y.}~\bibnamefont {Qiu}}, \bibinfo
		{author} {\bibfnamefont {V.}~\bibnamefont {Thampy}}, \bibinfo {author}
		{\bibfnamefont {A.~T.}\ \bibnamefont {Savici}}, \bibinfo {author}
		{\bibfnamefont {J.~A.}\ \bibnamefont {Rodriguez}}, \ and\ \bibinfo {author}
		{\bibfnamefont {C.}~\bibnamefont {Broholm}},\ }\href {\doibase
		10.1038/nmat2800} {\bibfield  {journal} {\bibinfo  {journal} {Nat. Mater.}\
		}\textbf {\bibinfo {volume} {9}},\ \bibinfo {pages} {718} (\bibinfo {year}
		{2010})}\BibitemShut {NoStop}%
	\bibitem [{\citenamefont {Bendele}\ \emph {et~al.}(2010)\citenamefont
		{Bendele}, \citenamefont {Babkevich}, \citenamefont {Katrych}, \citenamefont
		{Gvasaliya}, \citenamefont {Pomjakushina}, \citenamefont {Conder},
		\citenamefont {Roessli}, \citenamefont {Boothroyd}, \citenamefont
		{Khasanov},\ and\ \citenamefont {Keller}}]{Bendele2010}%
	\BibitemOpen
	\bibfield  {author} {\bibinfo {author} {\bibfnamefont {M.}~\bibnamefont
			{Bendele}}, \bibinfo {author} {\bibfnamefont {P.}~\bibnamefont {Babkevich}},
		\bibinfo {author} {\bibfnamefont {S.}~\bibnamefont {Katrych}}, \bibinfo
		{author} {\bibfnamefont {S.~N.}\ \bibnamefont {Gvasaliya}}, \bibinfo {author}
		{\bibfnamefont {E.}~\bibnamefont {Pomjakushina}}, \bibinfo {author}
		{\bibfnamefont {K.}~\bibnamefont {Conder}}, \bibinfo {author} {\bibfnamefont
			{B.}~\bibnamefont {Roessli}}, \bibinfo {author} {\bibfnamefont {A.~T.}\
			\bibnamefont {Boothroyd}}, \bibinfo {author} {\bibfnamefont {R.}~\bibnamefont
			{Khasanov}}, \ and\ \bibinfo {author} {\bibfnamefont {H.}~\bibnamefont
			{Keller}},\ }\href@noop {} {\bibfield  {journal} {\bibinfo  {journal} {Phys.
				Rev. B}\ }\textbf {\bibinfo {volume} {82}},\ \bibinfo {pages} {212504}
		(\bibinfo {year} {2010})}\BibitemShut {NoStop}%
	\bibitem [{\citenamefont {Rodriguez}\ \emph {et~al.}(2011)\citenamefont
		{Rodriguez}, \citenamefont {Stock}, \citenamefont {Hsieh}, \citenamefont
		{Butch}, \citenamefont {Paglione},\ and\ \citenamefont {Green}}]{rodr11}%
	\BibitemOpen
	\bibfield  {author} {\bibinfo {author} {\bibfnamefont {E.~E.}\ \bibnamefont
			{Rodriguez}}, \bibinfo {author} {\bibfnamefont {C.}~\bibnamefont {Stock}},
		\bibinfo {author} {\bibfnamefont {P.-Y.}\ \bibnamefont {Hsieh}}, \bibinfo
		{author} {\bibfnamefont {N.~P.}\ \bibnamefont {Butch}}, \bibinfo {author}
		{\bibfnamefont {J.}~\bibnamefont {Paglione}}, \ and\ \bibinfo {author}
		{\bibfnamefont {M.~A.}\ \bibnamefont {Green}},\ }\href {\doibase
		10.1039/C1SC00114K} {\bibfield  {journal} {\bibinfo  {journal} {Chem. Sci.}\
		}\textbf {\bibinfo {volume} {2}},\ \bibinfo {pages} {1782} (\bibinfo {year}
		{2011})}\BibitemShut {NoStop}%
	\bibitem [{\citenamefont {Dong}\ \emph {et~al.}(2011)\citenamefont {Dong},
		\citenamefont {Wang}, \citenamefont {Li}, \citenamefont {Chen}, \citenamefont
		{Yuan},\ and\ \citenamefont {Fang}}]{Dong2011}%
	\BibitemOpen
	\bibfield  {author} {\bibinfo {author} {\bibfnamefont {C.}~\bibnamefont
			{Dong}}, \bibinfo {author} {\bibfnamefont {H.}~\bibnamefont {Wang}}, \bibinfo
		{author} {\bibfnamefont {Z.}~\bibnamefont {Li}}, \bibinfo {author}
		{\bibfnamefont {J.}~\bibnamefont {Chen}}, \bibinfo {author} {\bibfnamefont
			{H.~Q.}\ \bibnamefont {Yuan}}, \ and\ \bibinfo {author} {\bibfnamefont
			{M.}~\bibnamefont {Fang}},\ }\href@noop {} {\bibfield  {journal} {\bibinfo
			{journal} {Phys. Rev. B}\ }\textbf {\bibinfo {volume} {84}},\ \bibinfo
		{pages} {224506} (\bibinfo {year} {2011})}\BibitemShut {NoStop}%
	\bibitem [{\citenamefont {Koshika}\ \emph {et~al.}(2013)\citenamefont
		{Koshika}, \citenamefont {Usui}, \citenamefont {Adachi}, \citenamefont
		{Watanabe}, \citenamefont {Sakano}, \citenamefont {Simayi},\ and\
		\citenamefont {Yoshizawa}}]{Koshika2013}%
	\BibitemOpen
	\bibfield  {author} {\bibinfo {author} {\bibfnamefont {Y.}~\bibnamefont
			{Koshika}}, \bibinfo {author} {\bibfnamefont {T.}~\bibnamefont {Usui}},
		\bibinfo {author} {\bibfnamefont {S.}~\bibnamefont {Adachi}}, \bibinfo
		{author} {\bibfnamefont {T.}~\bibnamefont {Watanabe}}, \bibinfo {author}
		{\bibfnamefont {K.}~\bibnamefont {Sakano}}, \bibinfo {author} {\bibfnamefont
			{S.}~\bibnamefont {Simayi}}, \ and\ \bibinfo {author} {\bibfnamefont
			{M.}~\bibnamefont {Yoshizawa}},\ }\href {\doibase 10.7566/JPSJ.82.023703}
	{\bibfield  {journal} {\bibinfo  {journal} {J. Phys. Soc. Jpn.}\ }\textbf
		{\bibinfo {volume} {82}},\ \bibinfo {pages} {023703} (\bibinfo {year}
		{2013})}\BibitemShut {NoStop}%
	\bibitem [{\citenamefont {Sun}\ \emph {et~al.}(2013)\citenamefont {Sun},
		\citenamefont {Tsuchiya}, \citenamefont {Yamada}, \citenamefont {Taen},
		\citenamefont {Pyon}, \citenamefont {Shi},\ and\ \citenamefont
		{Tamegai}}]{Sun2013}%
	\BibitemOpen
	\bibfield  {author} {\bibinfo {author} {\bibfnamefont {Y.}~\bibnamefont
			{Sun}}, \bibinfo {author} {\bibfnamefont {Y.}~\bibnamefont {Tsuchiya}},
		\bibinfo {author} {\bibfnamefont {T.}~\bibnamefont {Yamada}}, \bibinfo
		{author} {\bibfnamefont {T.}~\bibnamefont {Taen}}, \bibinfo {author}
		{\bibfnamefont {S.}~\bibnamefont {Pyon}}, \bibinfo {author} {\bibfnamefont
			{Z.}~\bibnamefont {Shi}}, \ and\ \bibinfo {author} {\bibfnamefont
			{T.}~\bibnamefont {Tamegai}},\ }\href {\doibase 10.7566/JPSJ.82.115002}
	{\bibfield  {journal} {\bibinfo  {journal} {J. Phys. Soc. Jpn.}\ }\textbf
		{\bibinfo {volume} {82}},\ \bibinfo {pages} {115002} (\bibinfo {year}
		{2013})}\BibitemShut {NoStop}%
	\bibitem [{\citenamefont {Zaliznyak}\ \emph
		{et~al.}(2015{\natexlab{a}})\citenamefont {Zaliznyak}, \citenamefont
		{Savici}, \citenamefont {Lumsden}, \citenamefont {Tsvelik}, \citenamefont
		{Hu},\ and\ \citenamefont {Petrovic}}]{zali15}%
	\BibitemOpen
	\bibfield  {author} {\bibinfo {author} {\bibfnamefont {I.}~\bibnamefont
			{Zaliznyak}}, \bibinfo {author} {\bibfnamefont {A.~T.}\ \bibnamefont
			{Savici}}, \bibinfo {author} {\bibfnamefont {M.}~\bibnamefont {Lumsden}},
		\bibinfo {author} {\bibfnamefont {A.}~\bibnamefont {Tsvelik}}, \bibinfo
		{author} {\bibfnamefont {R.}~\bibnamefont {Hu}}, \ and\ \bibinfo {author}
		{\bibfnamefont {C.}~\bibnamefont {Petrovic}},\ }\href {\doibase
		10.1073/pnas.1503559112} {\bibfield  {journal} {\bibinfo  {journal} {Proc.
				Natl. Acad. Sci. USA}\ }\textbf {\bibinfo {volume} {112}},\ \bibinfo {pages}
		{10316} (\bibinfo {year} {2015}{\natexlab{a}})}\BibitemShut {NoStop}%
	\bibitem [{\citenamefont {Jinsheng}\ \emph {et~al.}(2011)\citenamefont
		{Jinsheng}, \citenamefont {Guangyong}, \citenamefont {Genda}, \citenamefont
		{Tranquada},\ and\ \citenamefont {Birgeneau}}]{JWen2011}%
	\BibitemOpen
	\bibfield  {author} {\bibinfo {author} {\bibfnamefont {W.}~\bibnamefont
			{Jinsheng}}, \bibinfo {author} {\bibfnamefont {X.}~\bibnamefont {Guangyong}},
		\bibinfo {author} {\bibfnamefont {G.}~\bibnamefont {Genda}}, \bibinfo
		{author} {\bibfnamefont {J.~M.}\ \bibnamefont {Tranquada}}, \ and\ \bibinfo
		{author} {\bibfnamefont {R.~J.}\ \bibnamefont {Birgeneau}},\ }\href@noop {}
	{\bibfield  {journal} {\bibinfo  {journal} {Rep. Prog. Phys.}\ }\textbf
		{\bibinfo {volume} {74}},\ \bibinfo {pages} {124503} (\bibinfo {year}
		{2011})}\BibitemShut {NoStop}%
	\bibitem [{\citenamefont {Lynn}\ \emph {et~al.}(2012)\citenamefont {Lynn},
		\citenamefont {Chen}, \citenamefont {Chang}, \citenamefont {Zhao},
		\citenamefont {Chi}, \citenamefont {II}, \citenamefont {Ueland},\ and\
		\citenamefont {Erwin}}]{Lynn2012}%
	\BibitemOpen
	\bibfield  {author} {\bibinfo {author} {\bibfnamefont {J.~W.}\ \bibnamefont
			{Lynn}}, \bibinfo {author} {\bibfnamefont {Y.}~\bibnamefont {Chen}}, \bibinfo
		{author} {\bibfnamefont {S.}~\bibnamefont {Chang}}, \bibinfo {author}
		{\bibfnamefont {Y.}~\bibnamefont {Zhao}}, \bibinfo {author} {\bibfnamefont
			{S.}~\bibnamefont {Chi}}, \bibinfo {author} {\bibfnamefont {W.~R.}\
			\bibnamefont {II}}, \bibinfo {author} {\bibfnamefont {B.~G.}\ \bibnamefont
			{Ueland}}, \ and\ \bibinfo {author} {\bibfnamefont {R.~W.}\ \bibnamefont
			{Erwin}},\ }\href@noop {} {\bibfield  {journal} {\bibinfo  {journal} {Journal
				of Research of the National Institute of Standards and Technology}\ }\textbf
		{\bibinfo {volume} {117}},\ \bibinfo {pages} {61} (\bibinfo {year}
		{2012})}\BibitemShut {NoStop}%
	\bibitem [{\citenamefont {Wen}\ \emph {et~al.}(2009)\citenamefont {Wen},
		\citenamefont {Xu}, \citenamefont {Xu}, \citenamefont {Lin}, \citenamefont
		{Li}, \citenamefont {Ratcliff}, \citenamefont {Gu},\ and\ \citenamefont
		{Tranquada}}]{Wen2009}%
	\BibitemOpen
	\bibfield  {author} {\bibinfo {author} {\bibfnamefont {J.}~\bibnamefont
			{Wen}}, \bibinfo {author} {\bibfnamefont {G.}~\bibnamefont {Xu}}, \bibinfo
		{author} {\bibfnamefont {Z.}~\bibnamefont {Xu}}, \bibinfo {author}
		{\bibfnamefont {Z.~W.}\ \bibnamefont {Lin}}, \bibinfo {author} {\bibfnamefont
			{Q.}~\bibnamefont {Li}}, \bibinfo {author} {\bibfnamefont {W.}~\bibnamefont
			{Ratcliff}}, \bibinfo {author} {\bibfnamefont {G.}~\bibnamefont {Gu}}, \ and\
		\bibinfo {author} {\bibfnamefont {J.~M.}\ \bibnamefont {Tranquada}},\
	}\href@noop {} {\bibfield  {journal} {\bibinfo  {journal} {Phys. Rev. B}\
		}\textbf {\bibinfo {volume} {80}},\ \bibinfo {pages} {104506} (\bibinfo
		{year} {2009})}\BibitemShut {NoStop}%
	\bibitem [{\citenamefont {Aluru}\ \emph {et~al.}(2017)\citenamefont {Aluru},
		\citenamefont {Zhou}, \citenamefont {Essig}, \citenamefont {Reid},
		\citenamefont {Tsurkan}, \citenamefont {Loidl}, \citenamefont {Deisenhofer},\
		and\ \citenamefont {Wahl}}]{alur17}%
	\BibitemOpen
	\bibfield  {author} {\bibinfo {author} {\bibfnamefont {R.}~\bibnamefont
			{Aluru}}, \bibinfo {author} {\bibfnamefont {H.}~\bibnamefont {Zhou}},
		\bibinfo {author} {\bibfnamefont {A.}~\bibnamefont {Essig}}, \bibinfo
		{author} {\bibfnamefont {J.-P.}\ \bibnamefont {Reid}}, \bibinfo {author}
		{\bibfnamefont {V.}~\bibnamefont {Tsurkan}}, \bibinfo {author} {\bibfnamefont
			{A.}~\bibnamefont {Loidl}}, \bibinfo {author} {\bibfnamefont
			{J.}~\bibnamefont {Deisenhofer}}, \ and\ \bibinfo {author} {\bibfnamefont
			{P.}~\bibnamefont {Wahl}},\ }\href@noop {} {\enquote {\bibinfo {title}
			{{Atomic-scale coexistence of short-range magnetic order and
					superconductivity in Fe$_{1+y}$Se$_{0.1}$Te$_{0.9}$}},}\ } (\bibinfo {year}
	{2017}),\ \Eprint {http://arxiv.org/abs/https://arxiv.org/abs/1711.10389}
	{https://arxiv.org/abs/1711.10389} \BibitemShut {NoStop}%
	\bibitem [{\citenamefont {Zaliznyak}\ \emph
		{et~al.}(2015{\natexlab{b}})\citenamefont {Zaliznyak}, \citenamefont
		{Savici}, \citenamefont {Lumsden}, \citenamefont {Tsvelik}, \citenamefont
		{Hu},\ and\ \citenamefont {Petrovic}}]{Zaliznyak2015}%
	\BibitemOpen
	\bibfield  {author} {\bibinfo {author} {\bibfnamefont {I.}~\bibnamefont
			{Zaliznyak}}, \bibinfo {author} {\bibfnamefont {A.~T.}\ \bibnamefont
			{Savici}}, \bibinfo {author} {\bibfnamefont {M.}~\bibnamefont {Lumsden}},
		\bibinfo {author} {\bibfnamefont {A.}~\bibnamefont {Tsvelik}}, \bibinfo
		{author} {\bibfnamefont {R.}~\bibnamefont {Hu}}, \ and\ \bibinfo {author}
		{\bibfnamefont {C.}~\bibnamefont {Petrovic}},\ }\href@noop {} {\bibfield
		{journal} {\bibinfo  {journal} {Proc. Natl. Acad. Sci.}\ }\textbf {\bibinfo
			{volume} {112}},\ \bibinfo {pages} {10316} (\bibinfo {year}
		{2015}{\natexlab{b}})}\BibitemShut {NoStop}%
	\bibitem [{\citenamefont {Louca}\ \emph {et~al.}(2010)\citenamefont {Louca},
		\citenamefont {Horigane}, \citenamefont {Llobet}, \citenamefont {Arita},
		\citenamefont {Ji}, \citenamefont {Katayama}, \citenamefont {Konbu},
		\citenamefont {Nakamura}, \citenamefont {Koo}, \citenamefont {Tong},\ and\
		\citenamefont {Yamada}}]{louc10}%
	\BibitemOpen
	\bibfield  {author} {\bibinfo {author} {\bibfnamefont {D.}~\bibnamefont
			{Louca}}, \bibinfo {author} {\bibfnamefont {K.}~\bibnamefont {Horigane}},
		\bibinfo {author} {\bibfnamefont {A.}~\bibnamefont {Llobet}}, \bibinfo
		{author} {\bibfnamefont {R.}~\bibnamefont {Arita}}, \bibinfo {author}
		{\bibfnamefont {S.}~\bibnamefont {Ji}}, \bibinfo {author} {\bibfnamefont
			{N.}~\bibnamefont {Katayama}}, \bibinfo {author} {\bibfnamefont
			{S.}~\bibnamefont {Konbu}}, \bibinfo {author} {\bibfnamefont
			{K.}~\bibnamefont {Nakamura}}, \bibinfo {author} {\bibfnamefont {T.-Y.}\
			\bibnamefont {Koo}}, \bibinfo {author} {\bibfnamefont {P.}~\bibnamefont
			{Tong}}, \ and\ \bibinfo {author} {\bibfnamefont {K.}~\bibnamefont
			{Yamada}},\ }\href {\doibase 10.1103/PhysRevB.81.134524} {\bibfield
		{journal} {\bibinfo  {journal} {Phys. Rev. B}\ }\textbf {\bibinfo {volume}
			{81}},\ \bibinfo {pages} {134524} (\bibinfo {year} {2010})}\BibitemShut
	{NoStop}%
	\bibitem [{\citenamefont {Hu}\ \emph {et~al.}(2011)\citenamefont {Hu},
		\citenamefont {Zuo}, \citenamefont {Wen}, \citenamefont {Xu}, \citenamefont
		{Lin}, \citenamefont {Li}, \citenamefont {Gu}, \citenamefont {Park},\ and\
		\citenamefont {Greene}}]{hu11}%
	\BibitemOpen
	\bibfield  {author} {\bibinfo {author} {\bibfnamefont {H.}~\bibnamefont
			{Hu}}, \bibinfo {author} {\bibfnamefont {J.-M.}\ \bibnamefont {Zuo}},
		\bibinfo {author} {\bibfnamefont {J.}~\bibnamefont {Wen}}, \bibinfo {author}
		{\bibfnamefont {Z.}~\bibnamefont {Xu}}, \bibinfo {author} {\bibfnamefont
			{Z.}~\bibnamefont {Lin}}, \bibinfo {author} {\bibfnamefont {Q.}~\bibnamefont
			{Li}}, \bibinfo {author} {\bibfnamefont {G.}~\bibnamefont {Gu}}, \bibinfo
		{author} {\bibfnamefont {W.~K.}\ \bibnamefont {Park}}, \ and\ \bibinfo
		{author} {\bibfnamefont {L.~H.}\ \bibnamefont {Greene}},\ }\href@noop {}
	{\bibfield  {journal} {\bibinfo  {journal} {New J. Phys.}\ }\textbf {\bibinfo
			{volume} {13}},\ \bibinfo {pages} {053031} (\bibinfo {year}
		{2011})}\BibitemShut {NoStop}%
	\bibitem [{\citenamefont {Xu}\ \emph {et~al.}(2012)\citenamefont {Xu},
		\citenamefont {Wen}, \citenamefont {Zhao}, \citenamefont {Matsuda},
		\citenamefont {Ku}, \citenamefont {Liu}, \citenamefont {Gu}, \citenamefont
		{Lee}, \citenamefont {Birgeneau}, \citenamefont {Tranquada},\ and\
		\citenamefont {Xu}}]{xu12a}%
	\BibitemOpen
	\bibfield  {author} {\bibinfo {author} {\bibfnamefont {Z.}~\bibnamefont
			{Xu}}, \bibinfo {author} {\bibfnamefont {J.}~\bibnamefont {Wen}}, \bibinfo
		{author} {\bibfnamefont {Y.}~\bibnamefont {Zhao}}, \bibinfo {author}
		{\bibfnamefont {M.}~\bibnamefont {Matsuda}}, \bibinfo {author} {\bibfnamefont
			{W.}~\bibnamefont {Ku}}, \bibinfo {author} {\bibfnamefont {X.}~\bibnamefont
			{Liu}}, \bibinfo {author} {\bibfnamefont {G.}~\bibnamefont {Gu}}, \bibinfo
		{author} {\bibfnamefont {D.-H.}\ \bibnamefont {Lee}}, \bibinfo {author}
		{\bibfnamefont {R.~J.}\ \bibnamefont {Birgeneau}}, \bibinfo {author}
		{\bibfnamefont {J.~M.}\ \bibnamefont {Tranquada}}, \ and\ \bibinfo {author}
		{\bibfnamefont {G.}~\bibnamefont {Xu}},\ }\href@noop {} {\bibfield  {journal}
		{\bibinfo  {journal} {Phys. Rev. Lett.}\ }\textbf {\bibinfo {volume} {109}},\
		\bibinfo {pages} {227002} (\bibinfo {year} {2012})}\BibitemShut {NoStop}%
	\bibitem [{\citenamefont {Tsyrulin}\ \emph {et~al.}(2012)\citenamefont
		{Tsyrulin}, \citenamefont {Viennois}, \citenamefont {Giannini}, \citenamefont
		{Boehm}, \citenamefont {Jimenez-Ruiz}, \citenamefont {Omrani}, \citenamefont
		{Piazza},\ and\ \citenamefont {R{\o}nnow}}]{tsyr12}%
	\BibitemOpen
	\bibfield  {author} {\bibinfo {author} {\bibfnamefont {N.}~\bibnamefont
			{Tsyrulin}}, \bibinfo {author} {\bibfnamefont {R.}~\bibnamefont {Viennois}},
		\bibinfo {author} {\bibfnamefont {E.}~\bibnamefont {Giannini}}, \bibinfo
		{author} {\bibfnamefont {M.}~\bibnamefont {Boehm}}, \bibinfo {author}
		{\bibfnamefont {M.}~\bibnamefont {Jimenez-Ruiz}}, \bibinfo {author}
		{\bibfnamefont {A.~A.}\ \bibnamefont {Omrani}}, \bibinfo {author}
		{\bibfnamefont {B.~D.}\ \bibnamefont {Piazza}}, \ and\ \bibinfo {author}
		{\bibfnamefont {H.~M.}\ \bibnamefont {R{\o}nnow}},\ }\href@noop {} {\bibfield
		{journal} {\bibinfo  {journal} {New J. Phys.}\ }\textbf {\bibinfo {volume}
			{14}},\ \bibinfo {pages} {073025} (\bibinfo {year} {2012})}\BibitemShut
	{NoStop}%
	\bibitem [{\citenamefont {Yin}\ \emph {et~al.}(2011)\citenamefont {Yin},
		\citenamefont {Haule},\ and\ \citenamefont {Kotliar}}]{yin11b}%
	\BibitemOpen
	\bibfield  {author} {\bibinfo {author} {\bibfnamefont {Z.~P.}\ \bibnamefont
			{Yin}}, \bibinfo {author} {\bibfnamefont {K.}~\bibnamefont {Haule}}, \ and\
		\bibinfo {author} {\bibfnamefont {G.}~\bibnamefont {Kotliar}},\ }\href@noop
	{} {\bibfield  {journal} {\bibinfo  {journal} {Nat. Mater.}\ }\textbf
		{\bibinfo {volume} {10}},\ \bibinfo {pages} {932} (\bibinfo {year}
		{2011})}\BibitemShut {NoStop}%
	\bibitem [{\citenamefont {Yin}\ \emph {et~al.}(2010)\citenamefont {Yin},
		\citenamefont {Lee},\ and\ \citenamefont {Ku}}]{Yin2010}%
	\BibitemOpen
	\bibfield  {author} {\bibinfo {author} {\bibfnamefont {W.-G.}\ \bibnamefont
			{Yin}}, \bibinfo {author} {\bibfnamefont {C.-C.}\ \bibnamefont {Lee}}, \ and\
		\bibinfo {author} {\bibfnamefont {W.}~\bibnamefont {Ku}},\ }\href@noop {}
	{\bibfield  {journal} {\bibinfo  {journal} {Phys. Rev. Lett.}\ }\textbf
		{\bibinfo {volume} {105}},\ \bibinfo {pages} {107004} (\bibinfo {year}
		{2010})}\BibitemShut {NoStop}%
	\bibitem [{\citenamefont {Sarkar}\ \emph {et~al.}(2017)\citenamefont {Sarkar},
		\citenamefont {Van~Dyke}, \citenamefont {Sprau}, \citenamefont {Massee},
		\citenamefont {Welp}, \citenamefont {Kwok}, \citenamefont {Davis},\ and\
		\citenamefont {Morr}}]{sark17}%
	\BibitemOpen
	\bibfield  {author} {\bibinfo {author} {\bibfnamefont {S.}~\bibnamefont
			{Sarkar}}, \bibinfo {author} {\bibfnamefont {J.}~\bibnamefont {Van~Dyke}},
		\bibinfo {author} {\bibfnamefont {P.~O.}\ \bibnamefont {Sprau}}, \bibinfo
		{author} {\bibfnamefont {F.}~\bibnamefont {Massee}}, \bibinfo {author}
		{\bibfnamefont {U.}~\bibnamefont {Welp}}, \bibinfo {author} {\bibfnamefont
			{W.-K.}\ \bibnamefont {Kwok}}, \bibinfo {author} {\bibfnamefont {J.~C.~S.}\
			\bibnamefont {Davis}}, \ and\ \bibinfo {author} {\bibfnamefont {D.~K.}\
			\bibnamefont {Morr}},\ }\href {\doibase 10.1103/PhysRevB.96.060504}
	{\bibfield  {journal} {\bibinfo  {journal} {Phys. Rev. B}\ }\textbf {\bibinfo
			{volume} {96}},\ \bibinfo {pages} {060504} (\bibinfo {year}
		{2017})}\BibitemShut {NoStop}%
	\bibitem [{\citenamefont {Maier}\ \emph {et~al.}(2009)\citenamefont {Maier},
		\citenamefont {Graser}, \citenamefont {Scalapino},\ and\ \citenamefont
		{Hirschfeld}}]{maie09}%
	\BibitemOpen
	\bibfield  {author} {\bibinfo {author} {\bibfnamefont {T.~A.}\ \bibnamefont
			{Maier}}, \bibinfo {author} {\bibfnamefont {S.}~\bibnamefont {Graser}},
		\bibinfo {author} {\bibfnamefont {D.~J.}\ \bibnamefont {Scalapino}}, \ and\
		\bibinfo {author} {\bibfnamefont {P.}~\bibnamefont {Hirschfeld}},\
	}\href@noop {} {\bibfield  {journal} {\bibinfo  {journal} {Phys. Rev. B}\
		}\textbf {\bibinfo {volume} {79}},\ \bibinfo {pages} {134520} (\bibinfo
		{year} {2009})}\BibitemShut {NoStop}%
\end{thebibliography}

%

\end{document}